\documentclass[aps,pre,groupedaddress,showkeys,showpacs,amsmath]{revtex4}

\usepackage{amssymb}
\usepackage{amsmath}
\usepackage{amscd}
\usepackage{graphicx}
\usepackage{amsbsy}
\usepackage{color}

\begin{document}
\author{Achille Giacometti}
\affiliation{Dipartimento di Chimica Fisica, Universit\`a Ca' Foscari Venezia,
Calle Larga S. Marta DD2137, I-30123 Venezia, Italy}
\email{achille@unive.it}
\title{Effects of patch size and number within a simple model of patchy colloids}
\author{Fred Lado}
\affiliation{Department of Physics, North Carolina State University, Raleigh, North Carolina 
27695-8202, USA}
\email{fred_lado@ncsu.edu}
\author{Julio Largo}
\affiliation{Departamento de F\'isica Aplicada, Universidad de Cantabria, Avenida de los Castros
s/n, Santander 39005, Spain}
\author{Giorgio Pastore}
\affiliation{Dipartimento di Fisica dell' Universit\`a di Trieste 
and CNR-IOM UOS Democritos,
Strada Costiera 11, 34151 Trieste, Italy}
\email{pastore@ts.infn.it}
\author{Francesco Sciortino}
\affiliation{{Dipartimento di Fisica and
  CNR-ISC, Universit\`a di Roma {\em La Sapienza}, Piazzale A. Moro  2, 00185 Roma, Italy}}
\email{francesco.sciortino@uniroma1.it}

\date{\today}

\begin{abstract}
We report on a computer simulation  and integral equation study of a simple model of patchy spheres, each of whose surfaces is decorated with two opposite attractive caps, as a function of the fraction $\chi$ of covered attractive surface. The simple model explored --- the two-patch Kern-Frenkel model --- interpolates between a square-well and a hard-sphere potential on changing the coverage $\chi$. We show that integral equation theory provides quantitative predictions in the entire explored region of temperatures and densities from the square-well limit $\chi = 1.0$ down to $\chi \approx 0.6$. For smaller $\chi$, good numerical convergence of the equations is achieved only at temperatures larger than the gas-liquid critical point, where however integral equation theory provides a complete description of the angular dependence. These results are contrasted with those for the one-patch case. We investigate the remaining region 
of coverage via numerical simulation and show how the gas-liquid critical point moves to smaller densities and temperatures on decreasing  $\chi$. Below $\chi \approx 0.3$, crystallization prevents the possibility of observing the evolution of the line of critical points, providing the angular analog of the disappearance of the liquid as an equilibrium phase on decreasing the range for spherical potentials. Finally, we show that the stable ordered phase evolves on decreasing $\chi$ from a three-dimensional crystal of interconnected planes to a two-dimensional independent-planes structure to a one-dimensional fluid of  chains when the one-bond-per-patch limit is eventually reached. 
\end{abstract}


\maketitle
\section{Introduction}
\label{sec:intro}

Spherically symmetric potentials have become a well-established paradigm of colloidal science in past decades. \cite{Lyklema91} This is because, 
at a sufficiently coarse-grained level, colloidal surface composition can be regarded as uniform with a good degree of confidence, so that relevant 
interactions depend only on relative distances among the particles. Recent advances in chemical particle synthesis \cite{Manoharan03} 
have however challenged this view by emphasizing the fundamental role of surface colloidal heterogeneities and their detailed chemical 
compositions. This is particularly true for an important subclass of colloidal systems, namely proteins, where the presence of anisotropic 
interactions cannot be neglected, even at the minimal level. \cite{Lomakin99,McManus07,Liu07}
Directional interactions introduce novel properties in such systems. These properties depend both on the number of contacts ({\em i.e.}, 
the valency) and the amplitude of these interactions ({\em i.e.}, flexibility of the bonds), a notable example of this class being hydrogen-bond interactions, ubiquitous in biological, chemical, and physical processes. \cite{Robinson96,Jeffrey97} 

As a reasonable compromise between the high complexity of interactions governing the above systems and the necessary simplicity required for a minimal model, patchy-sphere models stand out for their remarkable success in this rapidly evolving field. \cite{Glotzer04,Glotzer07,Bianchi06,Wilber09} See Ref. \onlinecite{Pawar10} for a recent review on the subject.
 
Within this class of models, interactions are spread over a limited part of the surface, either concentrated over a number of pointlike spots 
\cite{Bianchi06,Bianchi07} or distributed over one or more extended regions. \cite{Chapman88,Kern03} While the former have the considerable advantage of a simple theoretical scheme \cite{Wertheim84} which allows a first semi-quantitative description, the latter can easily account for both the effect of the number of contacts and their amplitude, unlike ``spotty'' interactions which are always limited by the one-bond-per-site  constraint.

In this paper we consider a particular model due to Kern and Frenkel \cite{Kern03} of this patchy-spheres class wherein short-range attractive interactions --- of the square-well (SW) form --- are distributed over circular patches on otherwise hard spheres (HS). Interactions between particles (spheres) are then attractive in the SW-SW interfacial geometry or purely hard-sphere repulsive under the HS-SW or HS-HS interfacial geometries, and can sustain more than one bond --- in fact, as many as the geometry allows --- even in the case of a single patch assigned to each sphere. A number of real systems ranging from surfactants to globular proteins can be described with simplified interactions of these particular forms, with well-defined solvophilic and solvophobic regions, and despite their simplicity patchy hard spheres have already shown a remarkable richness of theoretical predictions. \cite{Chapman88,Kern03,Liu07,Foffi07,Fantoni07,Goegelein08} Notwithstanding the discontinous nature of the angular interactions, highly simplified integral equation approaches are possible, \cite{Fantoni07} but only very recently has a complete well-defined scheme, within the framework of the reference hypernetted-chain (RHNC) integral equation, been proposed and solved for patchy spheres. \cite{Giacometti09b} This integral equation belongs to a class of approximate closures which have been extensively exploited in the field of molecular associating fluids. \cite{Gray84} Its main advantage over other available approximations (other than its less-accurate parent HNC closure) lies in the fact that it relies on a single approximation, for the bridge function appearing in the exact relation between pair potential and pair distribution function $g(12)$, \cite{Hansen86,Gray84} to directly yield structural and thermodynamic properties that include the Helmholtz free energy and the chemical potential with no further approximations. \cite{Lado82a,Lado82b} In addition, it can be made to display enhanced consistency among different thermodynamic routes. \cite{Lado82} This is an important point when analyzing fluid-fluid phase diagrams such as we propose to do here. We thus build upon our previous work with the one-patch potential \cite{Giacometti09b} to study the two-patch case and its relationship with its one-patch counterpart. In addition to RHNC integral equation results, we provide dedicated Monte Carlo simulations which can assess the performance of RHNC. We find that RHNC provides a robust representation of both structural and thermophysical properties of the two-patch Kern-Frenkel model for a wide range of coverage $\chi$ (the ratio between attractive and total hard-sphere surface), extending from an isotropic SW to a bare HS potential. The competition arising between phase separation and polymerisation is discussed in terms of the angular dependence of the pair correlation function and the structure factor. Finally, a comparison between the one-patch and two-patch phase diagrams shows a strong impact on the different morphology and stable structures obtained in the two cases.

We also report numerical simulation results of the model in the region where the RHNC integral equations do not numerically converge, to explore the low temperature, small $\chi$ limit. We find that for $\chi <0.3$ it becomes impossible to investigate the low-temperature disordered phases, since the system quickly transforms into an ordered structure, which itself depends on the coverage value. Indeed, on decreasing $\chi$ one progressively enters the region where the maximum number of contacts per patch evolves from four to two and eventually reaches the one-bond-per-patch condition.  When three or four bonds per patch are possible, the observed ordered structure is a crystal of interconnected planes, while when only two contacts are possible, particles order themselves into a set of disconnected planes.

The patchy interaction model examined here can be regarded as a prototype of a special colloidal architecture where there exist competitive interactions on the colloidal surface that drive, by free energy minimization, the different colloidal particles through a spontaneous self-assembly process into complex superstructures whose final target can be experimentally probed and properly tuned. \cite{Walther09} The possibility, discussed in the present study, of identifying the position of the gas-liquid coexisting lines and its relative interplay with different structures, opens up fascinating scenarios in material science, on the possibility of novel material design exploiting a bottom-up process not requiring human intervention. 

\section{The two-patch Kern-Frenkel model}
\label{sec:model}

\begin{figure}[htbp]
\begin{center}
\vskip0.5cm
\includegraphics[width=14cm]{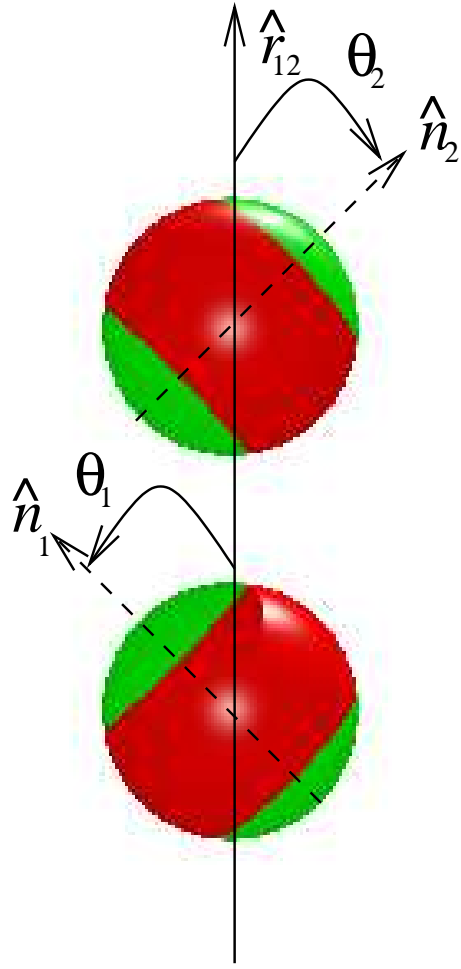}
\caption{The two-patch Kern-Frenkel potential. Each sphere is divided into an attractive
part (color code: green) and a repulsive part (color code: red). The attractive part is positioned
on two symmetrically distributed patches identified by unit vectors $\hat{\mathbf{n}}_{i}^{(t)}=\hat{\mathbf{n}}_{i}$ and $\hat{\mathbf{n}}_{i}^{(b)}=-\hat{\mathbf{n}}_{i}$ ($i=1,2$), where the orientation vectors $\hat{\mathbf{n}}_1$, $\hat{\mathbf{n}}_2$ define angles $\theta_{1}$, $\theta_{2}$ with the vector $\hat{\mathbf{r}}_{12}$ joining the centers of the two spheres and directed from sphere $1$ to sphere $2$. The particular case shown corresponds to a $40\%$ fraction of attractive surface (coverage $\chi$). 
\label{fig:fig1}}
\end{center}
\end{figure}

As a paradigmatic model for highly anisotropic interactions, we take the Kern-Frenkel \cite{Kern03} two-patch model where two attractive patches are symmetrically arranged as polar caps on a hard sphere of diameter $\sigma$. Each patch can be reckoned as the intersection of a spherical shell with a cone of semi-amplitude $\theta_0$ and vertex at the center of the sphere. Consider spheres $1$ and $2$ and let $\hat{\mathbf{r}}_{12}$ be the direction joining the two sphere centers, pointing from sphere $1$ to sphere $2$ (see Fig. \ref{fig:fig1}). The orientation of sphere $i$ is defined by a unit vector $\hat{\textbf{n}}_i \equiv \hat{\textbf{n}}_i^{(t)}$ passing outward through the center of one of its patches, to be arbitrarily designated as the ``top'' $(t)$ patch. The patch on the opposite, ``bottom'' $(b)$ pole is then identified with the outward normal $\hat{\textbf{n}}_i^{(b)}=-\hat{\textbf{n}}_i$.

Two spheres attract via a square-well potential of range $\lambda\sigma$ and depth $\epsilon$ if any combination of the two patches on each sphere are within a solid angle defined by $\theta_0$ and otherwise repel each other as hard spheres. The pair potential then reads \cite{Kern03}
\begin{eqnarray}
\Phi\left(12\right) &=& \phi \left(r_{12}\right) \Psi\left(\hat{\mathbf{n}}_1,
\hat{\mathbf{n}}_2,\hat{\mathbf{r}}_{12} \right),
\label{ie:eq1}
\end{eqnarray}
where
\begin{equation}
\phi\left(r\right)= \left\{ 
\begin{array}{ccc}
\infty,    &  &   0<r< \sigma    \\ 
- \epsilon, &  &   \sigma<r< \lambda \sigma   \\ 
0,          &  &   \lambda \sigma < r \ %
\end{array}%
\right.  \label{ie:eq2}
\end{equation}
and
\begin{equation}
\Psi\left(\hat{\mathbf{n}}_1,
\hat{\mathbf{n}}_2,\hat{\mathbf{r}}_{12}\right)= \left\{ 
\begin{array}{ccccc}
1,    & \text{if}  &   \hat{\mathbf{n}}_1^{\left(p_1\right)} \cdot \hat{\mathbf{r}}_{12} \ge \cos \theta_0 & \text{and} &  
-\hat{\mathbf{n}}_2^{\left(p_2\right)} \cdot \hat{\mathbf{r}}_{12} \ge \cos \theta_0 \\ 
0,    &  & &\text{otherwise} & \
\end{array}%
\right.  \label{ie:eq3}
\end{equation}
where $p_1,p_2=t$ or $b$ indicates which patch, top or bottom, is involved on each sphere. The unit vectors $\hat{\mathbf{n}}_{i}(\omega_{i})$ are defined by the 
spherical angles $\omega_i=(\theta_i,\varphi_i)$ in an arbitrarily oriented coordinate frame and $\hat{\mathbf{r}}_{12}(\Omega)$  is identified by the spherical 
angle $\Omega$ in the same frame. Reduced units, temperature $T^*=k_B T/\epsilon$ and density $\rho^*=\rho \sigma^3$, will be used throughout.

This model was introduced by Kern and Frenkel, \cite{Kern03} patterned after a similar model studied by Chapman \textit{et al.}, \cite{Chapman88} as a minimal model where both the distributions and the sizes of attractive surface regions on particles can be tuned. In this sense, the model constitutes a useful paradigm lying between spherically symmetric models which do not capture the specificity of surface groups, not even at the simplest possible level, and models with highly localized interactions having the single-bond, single-site limitation. \cite{Bianchi06,Bianchi07,Russo09} Several previous studies have already examined potentials of the Kern-Frenkel form using numerical simulations, \cite{Kern03,Liu07} corresponding-state arguments, \cite{Foffi07} highly simplified integral equation theories, \cite{Fantoni07} and pertubation theories. \cite{Goegelein08} More recently, \cite{Giacometti09b} the single patch Kern-Frenkel potential was studied using a more sophisticated integral  equation approach based on the RHNC  approximation coupled with rather precise and extensive Monte Carlo simulations. In the present paper, we extend this last study to the two-patch Kern-Frenkel potential and provide new methodologies specific for the angular distribution analysis.

We define the coverage $\chi$ as the fraction of the total sphere surface covered by attractive patches. Thus $\chi=1$ corresponds to a fully symmetric square-well potential while $\chi=0$ corresponds to a hard-sphere interaction and the model smoothly interpolates between these two extremes in the intermediate cases $0<\chi<1$. The two-patch potential is expected to present qualitative as well as quantitative differences with respect to its one-patch counterpart. One interesting question, for instance, concerns the subtle interplay between distribution and size of the attractive patches on the fluid-fluid phase separation diagram. It is now well established \cite{Kern03,Liu07,Fantoni07,Giacometti09b} that as coverage decreases the fluid-fluid coexistence line progressively diminishes in width and height. Indeed, this feature can be exploited to suppress phase separation altogether to enhance the possibility of studying glassy behavior \cite{Bianchi06,Bianchi07} and cannot be accounted for with a simple temperature and density rescaling, \cite{Fantoni07} although corresponding-state 
type of arguments can be proposed. \cite{Foffi07} On the other hand, the above mechanism can significantly depend on how the same reduced attractive region is distributed on the surface of particles. In Ref. \onlinecite{Fantoni07}, for instance, it was suggested that lines of decreasing critical temperature as a function of decreasing coverage, for the one-patch and two-patch Kern-Frenkel models with very short-range interactions, could cross each other at a specific 
coverage: for low coverages, critical temperatures for the one-patch model lie above the two-patch counterpart whereas the opposite is true for larger coverages. 
This would have far-reaching consequences on the phase diagram, as phase separation would occur at higher or lower temperatures for fixed coverage, depending on the specific allotment of the coverage. Another interesting issue regards micellization phenomena, present in the one-patch version of the model, \cite{Sciortino09} which is expected to be replaced by polymerization (or chaining) in the two-patch version. \cite{Sciortino07}
 
\section{Integral equation with RHNC closure and Monte Carlo simulations}
\label{sec: rhnc}

The Ornstein-Zernike (OZ) equation \cite{Hansen86} defines the direct correlation function $c(12)$ in terms of the pair correlation function $h(12)=g(12)-1$; it is convenient for computation to write it using the indirect correlation function $\gamma(12)=h(12)-c(12)$ instead of $h(12)$. We have then 
\begin{eqnarray}
\label{rhnc:eq1}
\gamma\left(12\right) &=& \frac{\rho}{4\pi} \int d\mathbf{r}_3 d\omega_3 \left[ \gamma\left(13\right)+c\left(13\right)\right] 
c\left(32\right).
\end{eqnarray}
A second, or ``closure,'' equation coupling $\gamma(12)$ and $c(12)$ is needed. The general form for this is \cite{Hansen86}
\begin{eqnarray}
\label{rhnc:eq2}
c\left(12\right)&=& \exp \left[-\beta \Phi\left(12\right)+\gamma\left(12\right)+
B\left(12\right)\right]-1-\gamma\left(12\right),
\end{eqnarray}
where $\beta=(k_BT)^{-1}$ and a third pair function, the so-called ``bridge'' function $B(12)$, has also been introduced. While known in a formal sense as a power series in density, \cite{Hansen86} $B(12)$ cannot in fact be evaluated exactly and at this point an approximation is unavoidable. The RHNC approximation replaces the unknown $B(12)$ with a known version $B_0(12)$ from some ``reference'' system. In practice, only the hard-sphere model is today well-enough known to play the role of reference system. Here we will use the Verlet-Weis-Henderson-Grundke parametrization \cite{Verlet72,Henderson75} for $B_0(12)=B_{\rm HS}(r_{12};\sigma_0)$, where $\sigma_0$ is the reference hard-sphere diameter. Some computational details of the RHNC integral equation approach can be found in Ref. \onlinecite{Giacometti09b} (see expecially Appendix A), so only the most relevant equations will be repeated here. 

Solution of the Ornstein-Zernike integral equation for molecular fluids \cite{Gray84} seemingly requires expansions in spherical harmonics of the angular dependence of all pair functions, a need that would be very problematic in the case of the discontinuous angular dependence in the present $\Phi\left(12\right)$. 
In fact, the integral equation algorithm allows $\Phi\left(12\right)$ to remain unexpanded. \cite{Giacometti09b} There is a potential problem however in evaluating the Gauss-Legendre quadratures used in the numerical solution, in that of the angles $\theta_1, \theta_2, \ldots, \theta_n$ used for an $n$th-order quadrature, none is likely to coincide with the angle $\theta_0$ defining the semi-amplitude of a patch. Thus the algorithm will not ``know'' the correct patch size. This problem is ameliorated in the following {\em ad hoc} fashion.

From the interaction $\Phi\left(12\right)$ of Eq. (\ref{ie:eq3}), the total coverage $\chi$ can be computed in terms of $\theta_0$ as
\begin{eqnarray}
\label{rhnc:eq3}
\chi^2 = \frac{1}{\left(4 \pi\right)^2}  \int &d \omega_1 d\omega_2&
\Bigl[\Theta\left(\cos \theta_1-\cos \theta_0\right) \Theta\left(-\cos \theta_2-\cos \theta_0\right) \\ \nonumber
& &
+\Theta\left(\cos \theta_1-\cos \theta_0\right) 
\Theta\left(\cos \theta_2-\cos \theta_0\right) \\ \nonumber
& &
+\Theta\left(-\cos \theta_1-\cos \theta_0\right) 
\Theta\left(-\cos \theta_2-\cos \theta_0\right) \\ \nonumber
& &
+\Theta\left(-\cos \theta_1-\cos \theta_0\right) 
\Theta\left(\cos \theta_2-\cos \theta_0\right) \Bigr],
\end{eqnarray}
where $\Theta(x)$ is the Heaviside step function, equal to $1$ if $x>0$ and $0$ if $x<0$. The integrals can be readily evaluated to give \cite{Kern03}
\begin{eqnarray}
\label{rhnc:eq5}
\chi&=& 2 \sin^2 \frac{\theta_0}{2}.
\end{eqnarray}
This quantity can also be numerically evaluated by Gauss-Legendre quadrature using the $n$ roots $\theta_j$ of the Legendre polynomial $P_n(\cos\theta)$ and the 
computed result compared with the exact value (\ref{rhnc:eq5}). We may then vary $n$ so as to find that number $n$ (typically kept between 30 and 40) that minimizes the known error in computing $\chi$. All Gaussian quadratures for that $\chi$ value are then evaluated with the same number $n$ of points, thus ensuring that minimal error arises from the selected angular grid.

In an axial {\bf r} frame \cite{Gray84} with $\hat{\mathbf{r}}_{12}=\hat{\mathbf{z}}$, the internal energy per particle in units $k_B T$ is obtained from 
\begin{eqnarray}
\label{rhnc:eq6}
\frac{\beta U}{N} &=&  -2\pi \rho \beta \epsilon \int_{\sigma}^{\lambda \sigma} dr ~ r^2
\left \langle g(r,\omega_1,\omega_2\right) \Psi\left(\omega_1, \omega_2 \right) \rangle_{\omega_1 \omega_2}, 
\end{eqnarray} 
where $\langle \ldots \rangle_{\omega}=(1/4\pi) \int d\omega \ldots $ denotes an average over spherical angle $\omega$ and where we have written out 
$g(12)=g(r,\omega_1,\omega_2)$. Similarly, the pressure $P$ is computed from the compressibility factor
\begin{eqnarray}
\label{rhnc:eq7}
\frac{\beta P}{\rho} &=& 1+ \frac{2}{3} \pi \rho \sigma^3 \left \{ 
\left \langle y\left(\sigma,\omega_1,\omega_2 \right) 
e^{\beta \epsilon \Psi\left(\omega_1,\omega_2\right)}\right \rangle_{\omega_1 \omega_2} 
 - \lambda^3 \left \langle y\left(\lambda \sigma,\omega_1,\omega_2 \right) \left[
e^{\beta \epsilon \Psi\left(\omega_1,\omega_2\right)}-1 \right] \right \rangle_{\omega_1
\omega_2} \right \},
\end{eqnarray}
where the cavity function $y(12)=g(12)e^{\beta \Phi(12)}$ has been introduced. The angular integrations in these expressions are evaluated with Gauss-Legendre and Gauss-Chebyshev quadratures. Finally, the dimensionless free energy per particle $\beta F/N$ and chemical potential $\beta\mu$ can also be directly computed from the pair functions produced by the RHNC equation; the overall calculation is optimized by choosing the reference hard sphere diameter $\sigma_0$ so as to minimize the free energy functional. \cite{Giacometti09b} We solve the RHNC equations numerically on $r$ and $k$ grids of $N_r=2048$ points, with intervals $\Delta r = 0.01 \sigma$ and $\Delta k = \pi/(N_r\Delta r)$, using a standard Picard iteration method. \cite{Hansen86} The square-well width is set at $\lambda=1.5$ as a reasonable value dictated by the availability of isotropic square well results. \cite{Giacometti09a} Further details of these and other computations can be found in Ref. \onlinecite{Giacometti09b}.

For an assessment of the performance of the RHNC integral equation, we also perform NVT, grand canonical, and Gibbs ensemble Monte Carlo (MC) simulations \cite{Frenkel02} following the path set in the one-patch case. \cite{Giacometti09b} Standard NVT MC simulations of a system of 1000 particles are used to compute structural information (pair correlation functions and structure factors) for comparison with integral equations results, whereas grand canonical and Gibbs ensemble MC (GEMC) are used to locate critical parameters and coexisting phases. The exact locations of the critical points (points connected by the thick dashed green line in Fig. \ref{fig:fig2}) have been obtained from the MC data assuming the Ising universality class and properly matching the density fluctuations with the known fluctuations of the magnetization close to the Ising critical point. \cite{Wilding97} For GEMC, we use a system of 1200 particles, which partition themselves into two boxes whose total volume is $4300\sigma^3$, corresponding to an average density of $\rho^*=0.27$. At the lowest temperature considered, this corresponds to roughly $1050$ particles in the liquid box and 150 particles in the gas box (of side $\approx 13\sigma$). On average, the code attempts one volume change every five particle-swap moves and 500 displacement moves. Each displacement move is composed of a simultaneous random translation of the particle center (uniformly distributed between $\pm 0.05 \sigma$) and a rotation (with an angle uniformly distributed between $\pm 0.1$ radians) around a random axis. We have studied systems of size $L=7$ up to $L=10$ to estimate the size dependence of the critical point, with an average of one insertion/deletion step every 500 displacement steps in the case of grand canonical Monte Carlo (GCMC). We have also performed a set of GCMC simulations for different choices of $T$ and $\mu$ to evaluate $\rho^*(\mu,T)$. See Ref. \onlinecite{Giacometti09b} and additional references therein for details.

\section{Numerical results}
\label{sec: numerical}
\subsection{Coexistence line}
\label{subsec:coex}

Locating coexistence lines is not an easy task within integral equation theory, given the fact that virtually all integral equations are unable to access the critical region with reliable precision due to significant thermodynamic inconsistencies among various possible routes to thermodynamics, a consequence of the approximation buried in the closure Eq. (\ref{rhnc:eq2}). The RHNC closure is no exception to this rule, but has the strong advantage of relying on a single approximation expressed by the choice of the reference bridge function $B_0(12)$, at odds with other available closures which require additional approximations in constructing various thermodynamic quantities such as the chemical potential. Here we follow the protocol outlined in Ref. \onlinecite{Giacometti09a} for the isotropic square-well potential and Ref. \onlinecite{Giacometti09b} for the one-patch Kern-Frenkel potential, where both the well-known pseudo-solutions shortcoming \cite{Belloni93} and the numerical drawbacks \cite{ElMendoub08} can be conveniently accounted for.

\begin{figure}[htbp]
\begin{center}
\vskip0.5cm
\includegraphics[width=14cm]{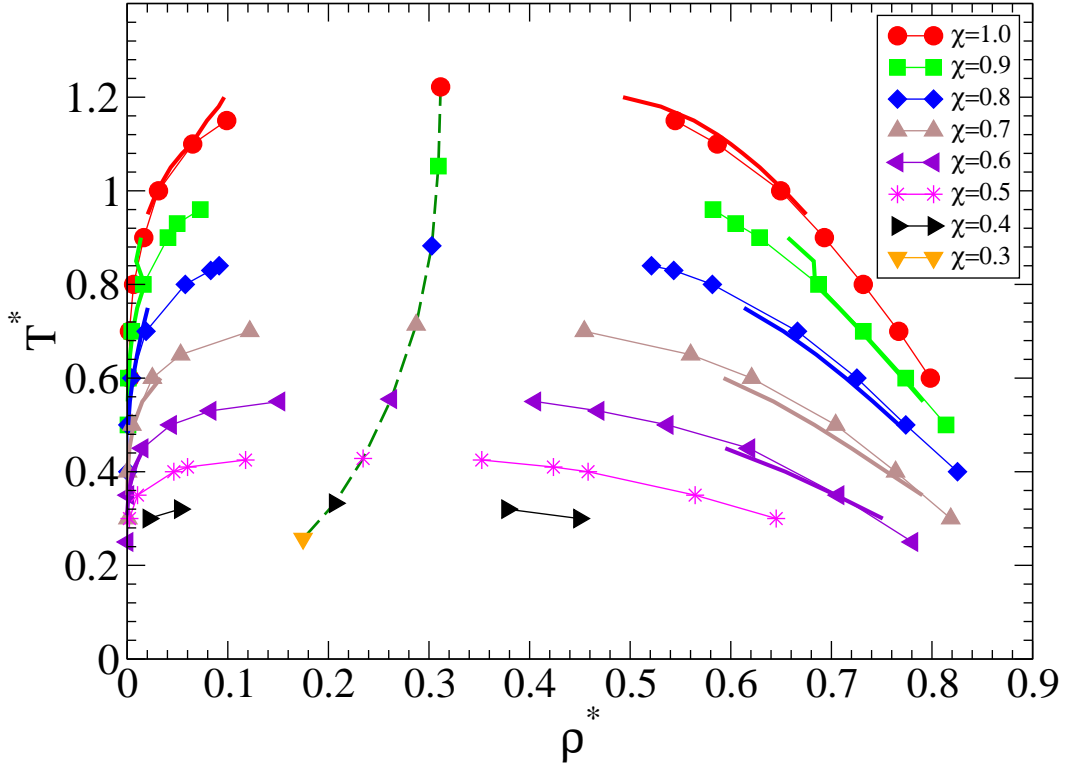}
\caption{Fluid-fluid coexistence lines of the two-patch model for different values of the coverage ranging from a full square-well potential down to $\chi=0.3$. Points represent MC results while thick solid lines report RHNC values (for $\chi \ge 0.6$). Thinner solid lines are a guide for the eye whereas the thick dashed line shows the  estimated GCMC critical point for any fixed coverage $\chi$. The MC data for $\chi=1.0$ (SW) coincide within the numerical error with the ones reported in Refs.~\onlinecite{vega92} and \onlinecite{thanosw}.
\label{fig:fig2}}
\end{center}
\end{figure}

Figure \ref{fig:fig2} depicts the location of the fluid-fluid coexistence line for the two-patch case upon varying the coverage $\chi$. The limiting case $\chi=1$ corresponds to the square-well potential. Both MC (points) and RHNC (thick solid lines) results are shown.  As previously noted, RHNC is not able to approach the critical point close enough to provide a direct estimate of its location. However, since it provides a quite good description of the low-temperature part of the coexistence line, we have tried to use these data to approximately locate the critical point. 

Visual inspection of the RHNC coexistence points reveals, in the cases where it is possible to go closer to the critical region, an unphysical change of curvature of the coexistence line moving from low to high temperature. For this reason, for each coverage we selected only data clearly consistent with a rectilinear diameter law. Then we fitted those data with the following function (corresponding to the first correction to the scaling \cite{Ley-KooGreen}):
\begin{equation}
\rho_l - \rho_g = a (T_c - T)^{\beta}(1 + b (T_c - T)^{\Delta}), \label{correction}
\end{equation}
where  the values of the exponents $\beta=0.325 $ and $\Delta=0.54$ are appropriate for the 3D Ising model universality class. \cite{LeveltSengers,ChenFisher} Once $T_c$ and the amplitudes $a$, $b$ have been determined, the critical density can be obtained from the rectilinear diameter best fit. The numerical results of such a procedure are compared with MC estimates of the critical points in Table \ref{tab:tab1}. It is evident that even though in general the validity of Eq. (\ref{correction}) is deemed to be limited to a smaller neighborhood of the critical point, \cite{LeveltSengers} in the present case it provides an acceptable procedure for a quick first estimate of the critical point location.

As the coverage decreases, the coexistence line shrinks and moves to lower temperature and density, as expected from an overall-decreasing attractive interaction. This trend can be tracked rather precisely by MC simulations down to remarkably low coverages ($\chi=0.3$) and RHNC correctly reproduces this evolution down to $\chi=0.6$ coverage. Below this value, more powerful algorithms are required to achieve good numerical convergence.

A few remarks are here in order. The coexistence curves shown in Fig. \ref{fig:fig2} are consistent with previous analogous results reported in 
Ref. \onlinecite{Kern03} but extend the range of temperatures and, more importantly, the range of coverages ($\chi$ values). This allows a quantitative measure
of the significant deviation from the simple mean-field-like results which can be obtained from the simple scaling (not shown) $T^{*} \to T^{*}/\chi$ as suggested by
the second-virial coefficient $B_2(T^{*})$ for this model, \cite{Kern03}
\begin{eqnarray}
\label{b2}
\frac{B_2\left(T^{*}\right)}{B_2^{\text{(HS)}}} &=& 1-\chi^2 \left(\lambda^3-1\right)\left(e^{\beta \epsilon}-1\right),
\end{eqnarray}
$B_2^{\text{(HS)}}$ being the hard-sphere result. This is also consistent with the breakdown of the above simple scaling at the level of the third virial coefficient,
derived in Ref. \onlinecite{Fantoni07} for the companion patchy sticky-hard-sphere model. 
As we shall see later on, the dependence of the critical temperature and density on \textit{both} the
coverage \textit{and} the number of patches is one of the main results of the present work. As a final point, we note that all curves in Fig. \ref{fig:fig2} collapse into a single master curve upon scaling $T^{*} \to T^{*}/ T_{c}^{*}$, in agreement with Ref. \onlinecite{Kern03}.

\subsection{Low-coverage results}
\label{subsec:low}

\begin{figure}[htbp]
\begin{center}
\vskip0.5cm
\includegraphics[width=14cm]{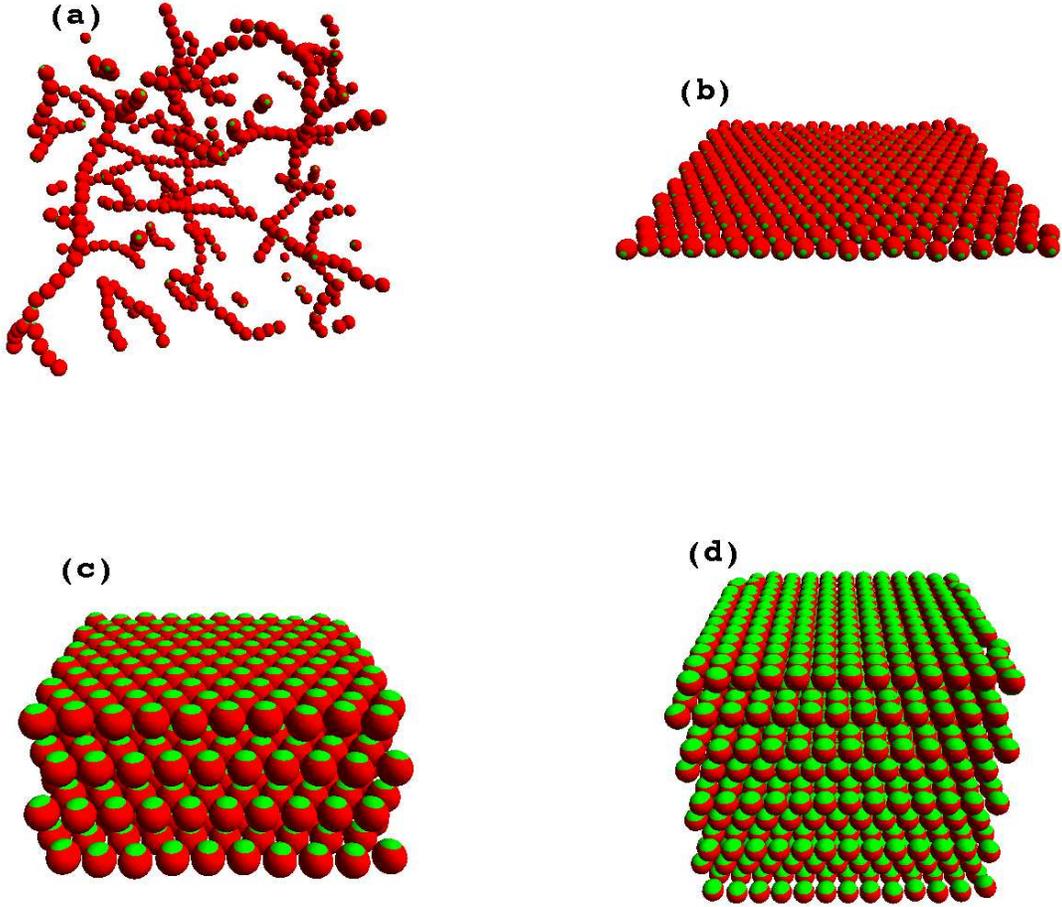}
\caption{Representation of the structures observed at small coverages $\chi$. (a) The case of coverages such that each patch can be involved in only one interaction. In this case, the system forms polydisperse chains. The snapshot here refers to the case $\rho^{*}=0.01$ and $T^{*}=0.07$. (b) 
Values of $\chi$ such that each patch can be involved in only two interactions. In this case, at low $T$ the system forms bonded planes interacting with each other only via excluded volume interactions. The snapshot shows one such plane. (c) Values of $\chi$ such that each patch can be involved in only three interactions. The crystal is now formed by interconnected planes, with a triangular arrangement of the particles in the plane.
Adjacent planes are shifted in such a way that each particle sits in correspondence to the center of a triangle of the previous and following planes. (d) Values of $\chi$ such that each patch can be involved in at most four interactions. The crystal is now formed by interconnected planes, with a square arrangement of the particles in the plane. Adjacent planes are shifted in such a way that each particle sits in correspondence to the center of a square of the previous and following planes.
\label{fig:fig3}}
\end{center}
\end{figure}

Below $\chi=0.3$, it becomes impossible to properly estimate the location of the critical point or the density of the coexisting gas and liquid phases. Indeed, the gas-liquid separation becomes pre-empted by crystallization into a structure that depends on the value of $\chi$.  Hence, the liquid phase, as an equilibrium phase, ceases to exist for small $\chi$.  This is strongly reminiscent of the disappearence of the liquid phase using spherical potentials when the range of the interaction becomes smaller than about 10\% of the particle diameter, \cite{Vliegenthart00,Pagan05,Liu05} thus providing the angular analog of the same phenomenon. Interestingly enough, the crystal structure which is spontaneously observed during the simulation depends on the value of $\chi$, since the $\chi$ value controls the maximum number of bonds per patch.  In the range of $\chi$ values such that each patch can be involved in four bonds, the observed ordered structure is made by planes exposing the SW parts to their surfaces (see Fig.~\ref{fig:fig3}(d)). Particles in the plane are located on a square lattice and adjacent planes are shifted in each direction by a half lattice constant, resulting in  a reduced energy of -4 per particle ({\em i.e.}, eight bonded neighbors). On decreasing $\chi$ below $\chi  \approx 0.118$, the region where only three bonds per particle are possible ($[\sqrt{3}(1+\Delta/\sigma)]^{-1} < \sin\theta_0 < [\sqrt{2}(1+\Delta/\sigma)]^{-1}$) is entered and  the crystal structure is made by interconnected planes of particles arranged in a triangular lattice (see Fig.~\ref{fig:fig3}(c)). For $[\sqrt{4}(1+\Delta/\sigma)]^{-1} < \sin\theta_0 < [\sqrt{3}(1+\Delta/\sigma)]^{-1}$ only two bonds per patch are possible and the system organizes into independent planes (see Fig.~\ref{fig:fig3}(b)), this time turning a HS surface to their neighboring planes. Particles in the plane are now arranged on a triangular lattice and  each patch is able to bind only to two different neighbors located in the same plane, resulting in a reduced energy per particle of $-2$. When $\sin\theta_0$ becomes smaller than the value $ [2(1+\Delta/\sigma)]^{-1}$ (corresponding to $\chi \approx 0.0572$), the one-bond-per-patch condition is reached and the system can form only isolated chains (see Fig.~\ref{fig:fig3}(a)). In this limit, the system is expected to behave as the  two single-bond-per-patch model. \cite{Sciortino07}
 
\subsection{Structural information}
\label{subsec:structural}

We turn our attention next to structural information, where the advantages of a reliable integral equation approach become evident. One has to keep in mind that Gibbs ensemble and GCMC simulations are particularly painstaking, due to the combined effect of the required low temperatures and the aggregation properties of the fluid (as detailed below), so that many of the state points examined here require several weeks of computer time. On the other hand, the RHNC integral equation, while rather demanding from an algorithmic point of view (see {\em e.g.} Appendix A of Ref. \onlinecite{Giacometti09b}) is a rapidly convergent scheme yielding solutions on the order of minutes, depending on the  temperatures considered. A more profound advantage stems from the fact that, within the approximation defined by the RHNC closure, all possible pair structural information is in fact exactly available, unlike MC calculations where, though available in principle, their statistics would be so limited as to make such calculations impractical. Thus, only the pair correlation function $g(12)$ averaged over angle $\hat{\mathbf{r}}_{12}(\Omega)$ is computed. By symmetry, the resulting pair function in this context depends only on $r\equiv r_{12}$ and 
$\cos\theta_{12}\equiv\hat{\mathbf{n}}_1 \cdot \hat{\mathbf{n}}_2$ and will be denoted here as $\bar{g}(r,\cos \theta_{12})$; see Appendix B in Ref. \onlinecite{Giacometti09b} for details.

Consider then the unaveraged pair correlation function $g(12)=g(r,\omega_1,\omega_2)$ from RHNC in an axial frame with $\hat{\mathbf{r}}_{12}=\hat{\mathbf{z}}$. Two noteworthy configurations occur when (a) all four patches lie along the same line (we denote this as the parallel $(||)$ configuration, with $\hat{\mathbf{n}}_1 \cdot \hat{\mathbf{n}}_2=\pm 1$, the actual labeling of each patch being unimportant) and when (b) patches on sphere $2$ lie on an axis perpendicular to those of sphere $1$ (in this case $\hat{\mathbf{n}}_1 \cdot \hat{\mathbf{n}}_2=0$, which we denote as the crossed (X) configuration). 

\begin{figure}[htbp]
\centering
\includegraphics[width=12cm]{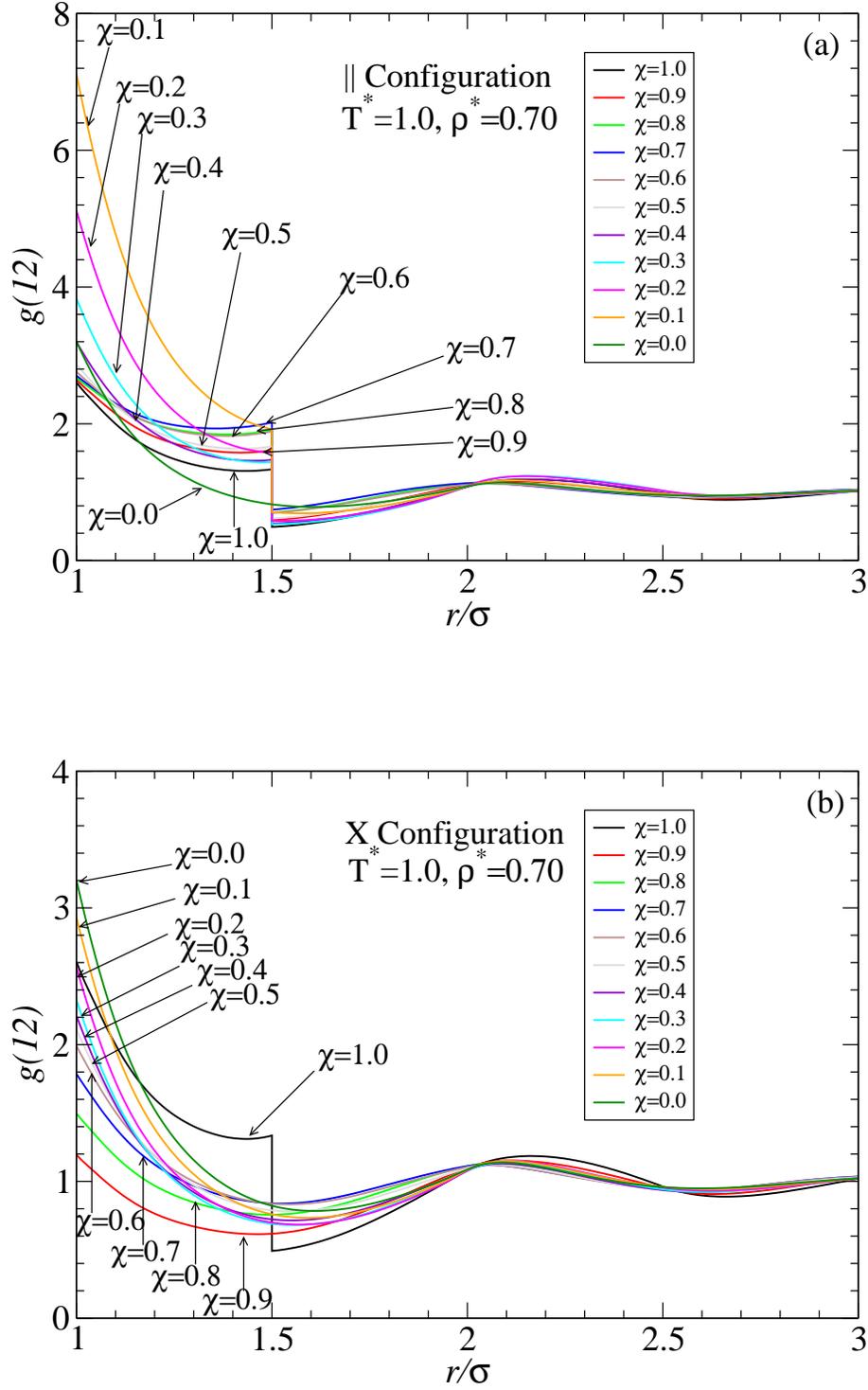}
\caption{Behavior of $g(12)$ from the RHNC equation for different coverages and two specific orientations of the patches: $||$ configuration corresponding to $\hat{\mathbf{n}}_1 \cdot \hat{\mathbf{n}}_2=\pm 1$ (a) and X configuration corresponding to $\hat{\mathbf{n}}_1 \cdot \hat{\mathbf{n}}_2=0$ (b). All curves are for a state point with $\lambda=1.5$, $\rho^{*}=0.7$, and $T^{*}=1.0$. Black and green lines show the limiting cases of square-well ($\chi=1$) and hard-sphere ($\chi=0$) potentials, respectively. Other coverages are $0.9-0.1$ for both the $||$ and X configurations.
\label{fig:fig4}}
\end{figure}

Figure \ref{fig:fig4} reports the results for the case $\lambda=1.5$, $\rho^{*}=0.7$, and $T^{*}=1.0$, which has been selected so that the fluid is above the coexistence line for all considered coverages, with  configurations $||$ and X in the top and bottom panels respectively. Values of coverages range from a full square-well potential ($\chi=1$) to a hard-sphere potential ($\chi=0$).

As coverage decreases, the contact value $r=\sigma^{+}$ of the $||$ configuration has the unusual behavior of first a slight increase from $\chi=1$ to $\chi=0.5$, followed by a more marked increase starting at $\chi=0.4$ up to the very small coverage $\chi=0.1$ limit which eventually backtracks to roughly the same value as at $\chi=0.4$ in the hard-sphere limit. At the opposite side of the well, $r=(\lambda \sigma)^{-}$, an even more erratic behavior is observed, with an increase in the range $0.7 < \chi < 1$, then a decrease for $0.3<\chi<0.6$, a new increase down to $\chi=0.1$, and a final sudden decrease to the hard sphere value $\chi=0$. A somewhat similar feature occurs in the X configuration where within the entire well region $\sigma^{+} \le r \le (\lambda \sigma)^{-}$ one observes a sudden decrease from $\chi=1$ to $\chi=0.9$ and a more gradual increase until reaching the highest value for the HS case. It is worth noting that in the X configuration there is no discontinuous jump at $r=\lambda \sigma$ for any value $\chi<1$. The reason for this has already been addressed in Ref. \onlinecite{Giacometti09b} for the one-patch case. Outside the first shell, there is a very weak dependence on the coverage, with a slight shift in the location of the second peak from a value of $r \approx 2.25 \sigma$ at the SW $\chi=1$ to a value of $r \approx 1.8 \sigma$ for lower $\chi$. 

\begin{figure}[htbp]
\centering
\includegraphics[width=12cm]{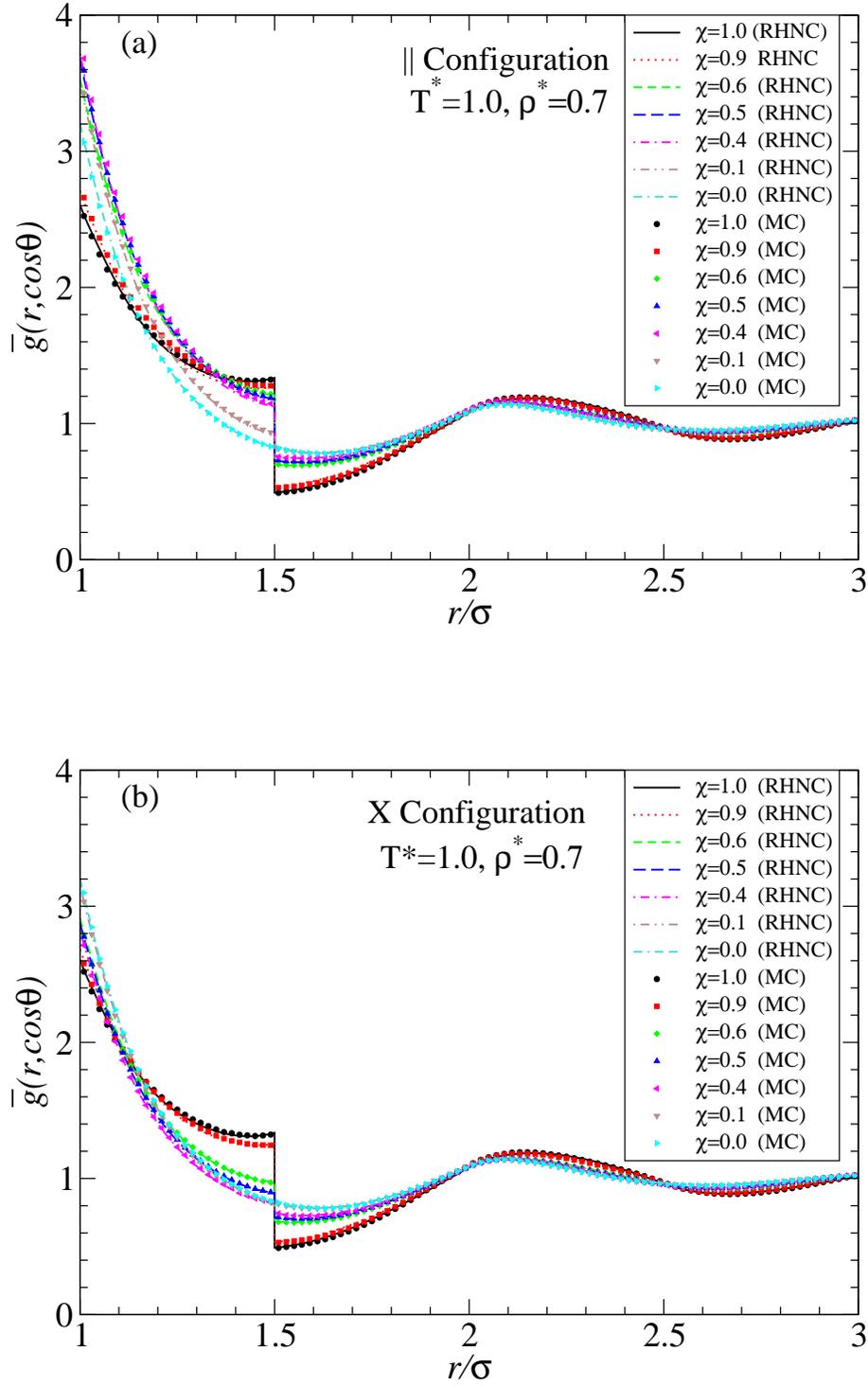}
\caption{Behavior of the averaged $\bar{g}(r,\cos \theta)$ for different coverages and two specific orientations of the patches: $||$ configuration corresponding to $\hat{\mathbf{n}}_1 \cdot \hat{\mathbf{n}}_2=\pm 1$ (a) and X configuration corresponding to $\hat{\mathbf{n}}_1 \cdot \hat{\mathbf{n}}_2=0$ (b). All curves are for a state point with $\lambda=1.5$, $\rho^{*}=0.7$, and $T^{*}=1.0$. Both RHNC and MC results are depicted for $\chi=0,0.1,0.4,0.5,0.6,0.9,1.0$.
\label{fig:fig5}}
\end{figure}

We compare RHNC integral equation results with MC simulations in Fig. \ref{fig:fig5}. As noted above, only the averaged pair function $\bar{g}(r,\cos \theta_{12})$ can be compared and this is done in the figure for different values of the coverage $\chi$ at the same state point ($T^{*}=1.0,\rho^{*}=0.7)$ and for the same $||$ and X configurations considered earlier. The good overall performance of RHNC in representing MC results is apparent as both contact values at the well edges and the jump discontinuities are very well reproduced. It is instructive to contrast these results with those of Fig. \ref{fig:fig4}, as many of the abrupt changes appearing in the actual pair correlation function are smoothed out by the orientational average carried out here. For instance, the characteristic jump at $r=\lambda \sigma$ of the $||$ configuration (top panel) progressively decreases as coverage is reduced and disappears in the hard-sphere limit. Conversely, the jump is present also in the X configuration (bottom panel) unlike the corresponding case of the full $g(12)$. In addition, the strong increase of the $||$ configuration for low coverages is not present in this figure; as remarked earlier, this level of detail in $g(12)$ is one of the main advantages of an integral equation approach. 

An additional useful quantity to consider, in view of its direct experimental access through scattering experiments, \cite{Hansen86} is the structure factor, which will be denoted $S_{000}(k)$ within our theoretical framework. \cite{Gray84,Lado82a,Giacometti09b} This is also strongly related, via Hankel transforms, to the radial distribution function $g_{000}(r)$, which is $g(12)$ averaged over all orientations $\omega_1$ and $\omega_2$ of the patches and of the relative angular position $\Omega$. Note that $g_{000}(r)$ is also the simplest rotational invariant (see Ref. \onlinecite{Gray84} and Appendix \ref{app:appb}).

The structure factor and the radial distribution function are reported in Fig. \ref{fig:fig6} for two representative values of coverage, $\chi=0.8$ and $\chi=0.2$, corresponding to almost fully attractive and almost fully repulsive limits. These values have been selected at the same state point previously considered ($T^{*}=1.0$, $\rho^{*}=0.7$) as having a very different behavior within the first shell $\sigma \le r \le \lambda \sigma$. This high-density result is also contrasted with a low-density state point $\rho^{*}=0.1$ at the same temperature, a value which, in the temperature-density plane, lies symmetrically with respect to the coexistence curves in the single fluid phase for all coverages (see Fig. \ref{fig:fig2}).

A few features are worth noting. For density $\rho^{*}=0.7$ there is a significant coverage dependence, where the contact value $g_{000}(\sigma^{+})$
for $\chi=0.2$ coverage is larger than that for the corresponding $\chi=0.8$ case and, conversely, the jump present at the other extreme $\lambda \sigma^{-}$ is much smaller in the former than in the latter case. A similar feature also occurs for the low-density state point $\rho^{*}=0.1$. This results from an angular average of the results given in Fig. \ref{fig:fig4}. Likewise, there is a marked difference in the behavior of the structure factor $S_{000}(k)$ for the high-density case $\rho^{*}=0.7$, both in the height of the first peak (related to the $g_{000}(\sigma^{+})$ value) and of the secondary peaks (related to the behaviors of $g_{000}(r)$ in the $\sigma < r < \lambda \sigma$ region and  of the $g_{000}(\lambda \sigma^{\pm})$ discontinuity). Similarly, in the low-density branch $\rho^{*}=0.1$, the large $S_{000}(0)$ value for the $\chi=0.8$ coverage case is signaling the approach to a spinodal instability which is clearly not present in the corresponding $\chi=0.2$ coverage.

One natural interpretation of the above results is the progressive rearrangement of the distribution within the first shell upon varying both the coverage and the density. To support this view, we consider the angular distribution within the first shell in the next subsection.

\subsection{Angular distribution}
\label{subsec:angular_dist2P}

The nonmonotonic dependence of $g(12)$ in terms of the distance $r/\sigma$ for decreasing coverage $\chi$, as illustrated in Fig. \ref{fig:fig4},  
is rather intriguing and requires an explanation. A similar, albeit different, feature occurs even in the one-patch case, as shown in Ref. \onlinecite{Giacometti09b}. We have tackled this in two ways, illustrated in the following.

All previous representations of $g(12)$ have been depicted in the molecular axial frame, where $\hat{\mathbf{n}}_1 \cdot \hat{\mathbf{r}}_{12} =1$, so that patches on sphere $1$ are parallel to the vector $\mathbf{r}_{12}$ joining sphere $1$ with sphere $2$. This is clearly preventing an understanding of the angular distribution of the patches around a given sphere $1$, that is, as a function of $\hat{\mathbf{r}}_{12}(\Omega)$.

\begin{figure}[htbp]
\centering
\includegraphics[width=14cm]{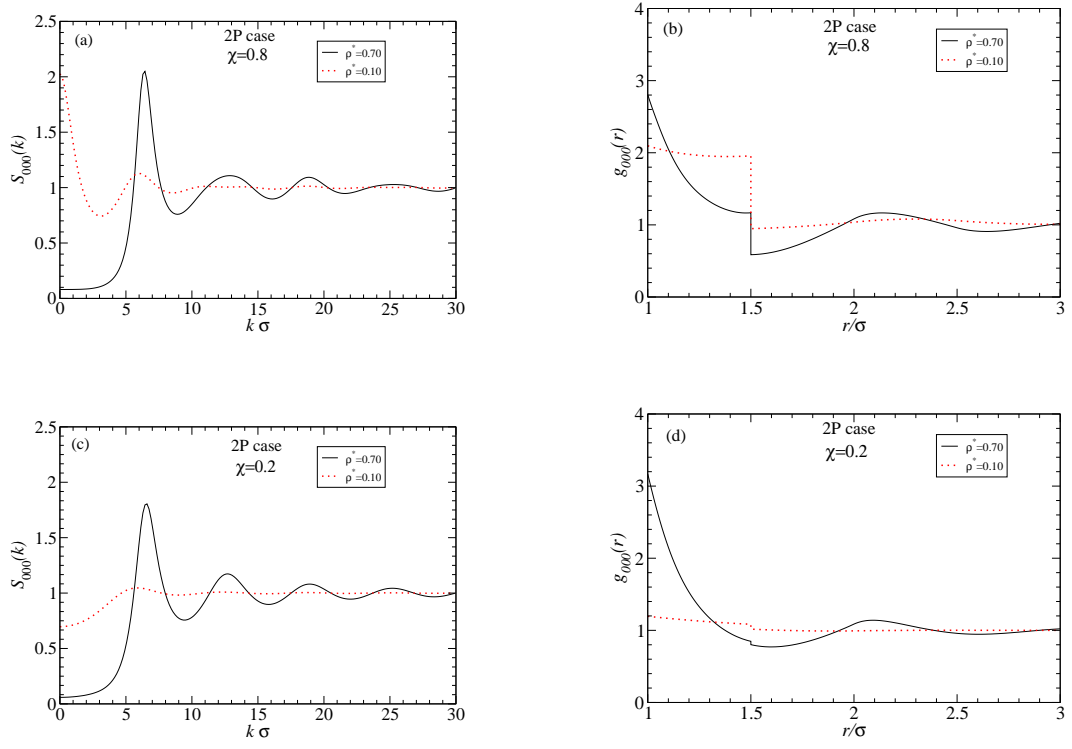}
\caption{Behavior of the structure factor $S_{000}(k)$ ((a) and (c) ) and the radial distribution function $g_{000}(r)$ ((b) and (d)) for coverages 
$\chi=0.8$ and $\chi=0.2$, respectively. Here $T^{*}=1.0$ and $\lambda=1.5$ as before, while densities are $\rho^{*}=0.7$ and $\rho^{*}=0.1$.
\label{fig:fig6}}
\end{figure}

This is however a needless restriction, as one can start from the expression for $g(12)$ in a general (laboratory) frame and study the dependence on the
angle $\Omega$ for fixed patch directions $\hat{\mathbf{n}}_1$ and $\hat{\mathbf{n}}_2$. In Figs. \ref{fig:fig4}, \ref{fig:fig5} and \ref{fig:fig6}, we notice that the main dependence on coverage $\chi$ stems from the region within the well, $\sigma \le r \le \lambda \sigma$; it sufficies therefore to investigate the \textit{average} $\Omega$ dependence by integrating over the radial variable $r$ within this region. We further note that there is azimuthal symmetry with respect to the $\varphi$ variable, so that we can focus on the $\theta$ dependence. The details of the analysis are reported in Appendix \ref{app:appa}, where it is shown that the relevant quantity is $\bar{g}(\theta,\theta_2)$, which is a function of the angle $\theta$ (polar dependence of $\hat{\mathbf{r}}_{12}(\Omega)$) and of the polar angle $\theta_2$ of the patches on particle $2$, given that the patches on particle $1$ lie along the $\hat{\mathbf{z}}$ axis. We report comparative calculations for both low ($\rho^{*}=0.1$) and high ($\rho^{*}=0.7$) density at identical temperature $T^{*}=1.0$ at two representative coverages, $\chi=0.8$, representing a case with almost all attraction, and $\chi=0.2$, as representative of an almost hard-sphere case. These are the same conditions considered in Fig. \ref{fig:fig6}; the results are reported in Fig. \ref{fig:fig7}. Let us consider first the high-density, $\rho^{*}=0.7$, situation as depicted in the two left panels for $\chi=0.8$ (a) and $\chi=0.2$ (c). Here, the $\chi=0.8$ case yields a very well-defined pattern with a periodically modulated distribution of the patches in symmetrical fashion as indicated by the trimodal distribution as a function of the relative positional angle $\theta$, so that $0,\pi/2,\pi$ are almost equally represented. (Note that $\theta=0,\pi$ are necessarily equivalent due to the up-down symmetry of the two-patch distribution.) The two interstitial minima are a consequence of the reduced valency --- the corresponding fully symmetrical result under this condition would be a flat distribution around the value $1.66$ in between the two maxima and minima --- so this slightly favours perpendicular orientation of the patches along the forward (or backward) direction and parallel orientation along the transversal direction. Under low coverage ($\chi=0.2$) conditions, on the other hand, there is clear evidence of a parallel orientation of the patches along the forward (or backward) direction, the opposite being true for a perpendicular orientation of the patches. This confirms the tendency to filament formations previously alluded to. The situation is even more evident at low density, $\rho^{*}=0.1$, as shown in the right two panels (b) and (d).  

\subsection{Coefficients of rotational invariants}
\label{subsec:rot_inv_coeff2P} 

Additional insights on the angular correlations of patch distributions can be obtained by considering other coefficients
$h^{l_{1} l_{2} l}(r)=g^{l_{1} l_{2} l}(r)-\delta_{{l}_{1}{l}_{2}{l},000} $ of the rotational invariants $\psi^{l_{1} l_{2} l}(r)$  
as defined in Eqs. (\ref{appa:eq1}) and (\ref{appa:eq2}); they have proven to be of invaluable help in discriminating between parallel and antiparallel configurations occurring in different models such as dipolar hard spheres \cite{Ganzenmuller07,Weis93} and Heisenberg spin fluids. \cite{Lado98} Some relevant properties of these coefficients are also listed in Appendix \ref{app:appb}, where we display explicit expressions for the first few coefficients.

In the two-patch case, we note that all coefficients with $l_1$ or $l_2$ odd vanish, so we depict the first nonvanishing coefficients 
$h^{220}(r)$, $h^{222}(r)$, $h^{022}(r)=h^{202}(r)$ in Fig. \ref{fig:fig8} for the same state points as before. Note that the left-most two curves, (a) and (c), correspond to density $\rho^{*}=0.7$, temperature $T^{*}=1.0$, coverages $\chi=0.8$ (a) and $\chi=0.2$ (c), and are plotted on the same scale. While qualitative trends are similar, the two cases have significantly different behavior. Within a given coverange $h^{220}(r)$ and $h^{222}(r)$ are almost coincident, with positive correlation in the well region $\sigma < r < \lambda r$, whereas $h^{022}(r)$ has decreasing positive correlation in the same region and negative correlation for $r> \lambda \sigma$. Numerical values, on the other hand, differ among each other, with small values for $h^{220}(r)$ and $h^{222}(r)$ in the high coverage case $\chi=0.8$ and significantly higher values in the low coverage limit $\chi=0.2$. 

Likewise, for the low-density state point $\rho^{*}=0.1$, $T^{*}=1.0$, right-hand-side plots (b) and (d) can be unambiguously discriminated between high (b) and low (d) coverages. We shall return to this point in the comparison with the one-patch results, where the physical meaning of these coefficients will be discussed.

\section{Comparison with one-patch results}
\label{sec:comparison}
\subsection{Phase diagram}
\label{subsec:phase_diagram}

Figure~\ref{fig:fig9} shows the critical parameters for the case of particles with one and two patches, both reported as a function of the total coverage. Here only MC results are reported in view of their precision and reliability. The $\chi=1$ limit corresponds to the SW case and coincides for both models. An analogous figure, for the case of adhesive patchy spheres (the limit of the present model for vanishing ranges), has been reported in Ref. \onlinecite{Fantoni07} 
within a simplified integral equation scheme.

With respect to the adhesive limit, the range of coverages which can be explored numerically is significantly wider (for both one-patch and two-patch cases). In the case of two patches, crystallization pre-empts the possibility of exploring the smaller values. In the one-patch case, the process of micelle formation, also observed experimentally, \cite{granick} suppresses the phase-separation process \cite{Sciortino09} at small $\chi$ values.    

The critical parameters decrease on decreasing $\chi$ for both one-patch and two-patch models. The behavior of  $T_{\rm c}$ can be explained on the basis of a progressive reduction of the attractive surface. The decrease in the critical density becomes significantly pronounced only for the smallest $\chi$ values and can be attributed to the lower local density requested for extensive bonding. Such behavior is analogous to the suppression of the critical density observed when the particle valence decreases. \cite{zacca1,Bianchi06} In the $\chi$ region where it is possible to evaluate the critical parameters, $T_{\rm c}$ and $\rho_{\rm c}$ for the two-patch case are always larger than the corresponding one-patch values, a trend which can be tentatively rationalized on the basis of the ability to form a larger number of contacts and higher local bonded densities for the case in which both poles of the particles can interact attractively with their neighbors. No evidence of a crossing between the two geometries is observed. Such a crossing has been predicted by  a theoretical approach based on a virial expansion up to third order in density and appropriate closures of the direct correlation function. \cite{Fantoni07} 

It is worth emphasizing that the above dependence on the number of patches, at a given coverage, provides clear evidence of the impossibility of rationalizing the change of the critical line on the basis of a trivial decrease of the attractive strength of interactions due to the reduction in coverage,
as alluded to in Section \ref{subsec:coex}.

For the sake of completness, we also report RHNC integral equation results for the most relevant thermodynamic quantities, as a function of the coverage $\chi$. These are shown in Table \ref{tab:tab2} for the same high-density state, $\rho^{*}=0.7$ at $T^{*}=1.0$, considered above for structural information. Here we present the reduced internal energy per particle and the reduced excess free energy per particle, $U /N \epsilon$ and $\beta F_{\rm{ex}}/N$, respectively, the reduced chemical potential $\beta \mu$, the compressibility factor $\beta P /\rho$, and the inverse compressibility $\beta \left(\partial P/\partial \rho\right)_T$. These results may be compared with those of Table IV in Ref. \onlinecite{Giacometti09b} listing the same quantities for the one-patch counterpart. (We ignore the tiny difference in densities between the two calculated states.) The last two columns give the reference HS diameter $\sigma_0$ stemming from the variational RHNC scheme (see Ref. \onlinecite{Giacometti09b} for details) and the average coordination number within the wells $\bar{z}$, whose one-patch counterparts are included in Tables IV and V of Ref. \onlinecite{Giacometti09b}, respectively. Note that $\bar{z}$ here is systematically larger than in the one-patch case, in qualitative agreement with the MC results of Fig.~\ref{fig:fig9}. 
  
\subsection{Angular distribution and coefficients of rotational invariants}
\label{subsec:angular_dist1P}

\begin{figure}[htbp]
\centering
\includegraphics[width=14cm]{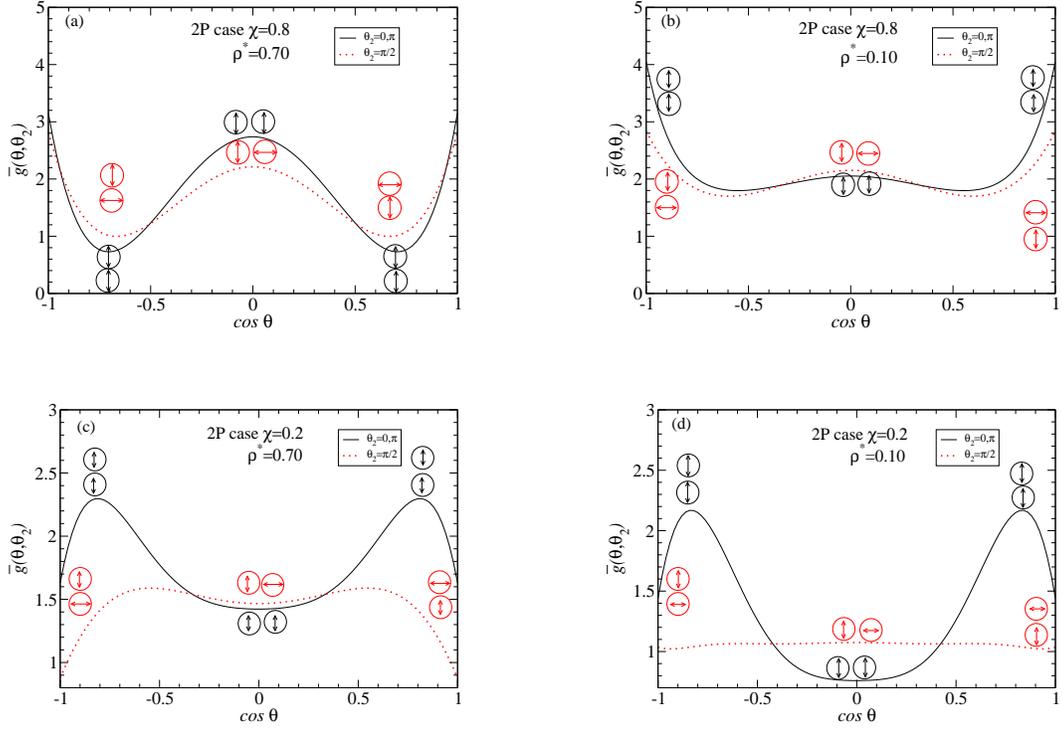}
\caption{Angular distributions $\bar{g}(\theta,\theta_2)$ as functions of $\cos \theta$ for two different orientations of the patches on sphere $2$, given that sphere $1$ is fixed with patches along the $\hat{\mathbf{z}}$ axis. Results are reported for two different coverages, $\chi=0.8$ ((a) and (b)) and $\chi=0.2$ ((c) and (d)), and two different densities, $\rho^{*}=0.7$ ((a) and (c)) and $\rho^{*}=0.1$ ((b) and (d)), at the same temperature $T^{*}=1.0$. Again, the square-well width was set to $\lambda=1.5$. The colored arrows are cartoons of the orientations of sphere $2$ patches, corresponding to $\theta_2=0,\pi/2$. Note that these are the same state points considered in Fig. \ref{fig:fig6}. 
\label{fig:fig7}}
\end{figure}

\begin{figure}[htbp]
\centering
\includegraphics[width=14cm]{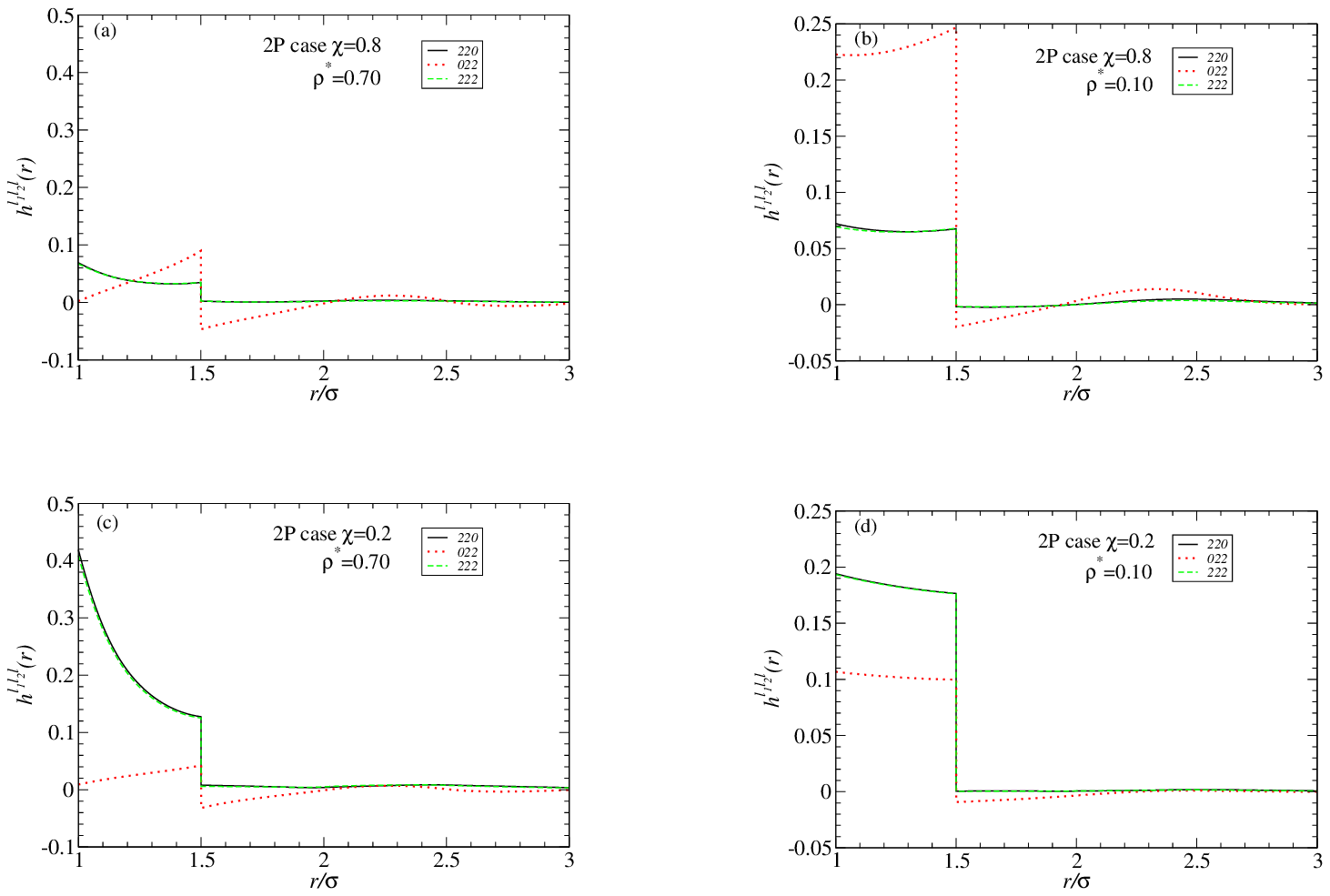}
\caption{Plot of rotational invariants  $h^{220}(r)$, $h^{022}(r)$, and $h^{222}(r)$ as functions of $r/\sigma$. Results are reported for two different coverages, $\chi=0.8$ ((a) and (b) ) and $\chi=0.2$ ((c) and (d)), and two different densities, $\rho^{*}=0.7$ ((a) and (c)) and $\rho^{*}=0.1$ ((b) and (d)), at the same temperature $T^{*}=1.0$ and square-well width $\lambda=1.5$. Again, these are the same state points considered in Figs. \ref{fig:fig6} and \ref{fig:fig7}. 
\label{fig:fig8}}
\end{figure}

\begin{figure}[htbp]
\begin{center}
\vskip0.5cm
\includegraphics[width=14cm]{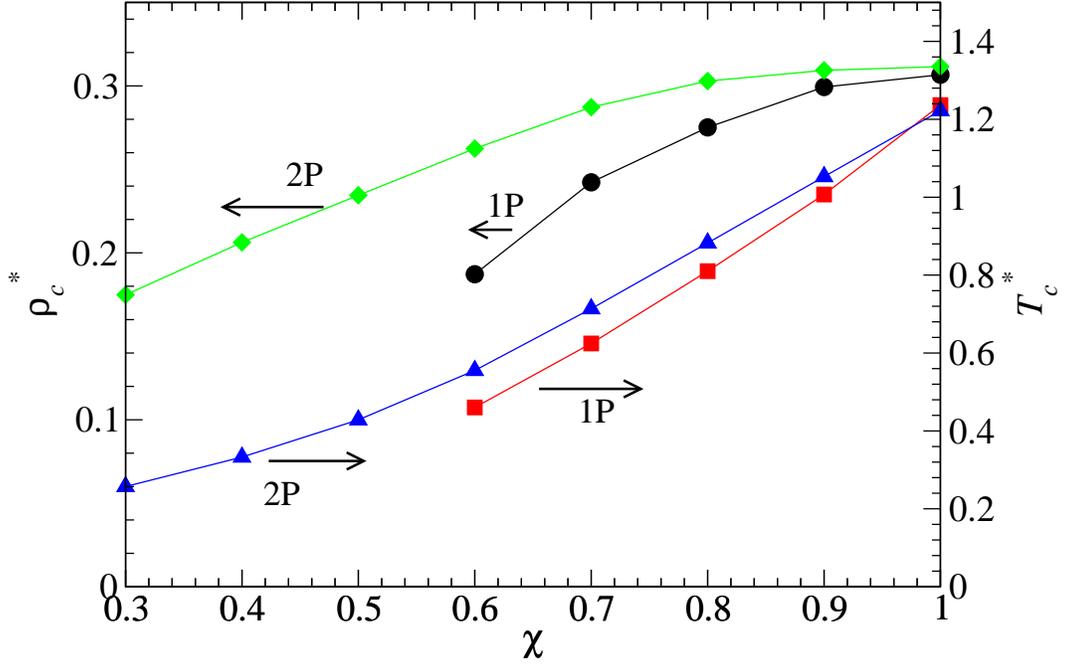}
\caption{Comparison between the critical parameters observed for the one-patch case (from Ref.~\onlinecite{januslungo}) and the two-patch case (this work).  
\label{fig:fig9}}
\end{center}
\end{figure}

Within the RHNC integral approach, the analysis of the angular distribution of patches within the first shell given in Section \ref{subsec:angular_dist2P} revealed that the cylindrical symmetry of a pair of patches (2P case) on each particle was very effective in driving the system to morphologically different configurations in the low ($20\%$) and high ($80\%$) coverage limits, as illustrated in Fig. \ref{fig:fig7}. It is natural to expect a very different situation in the single-patch case (1P case). This is indeed the case as further elaborated below.  

\begin{figure}[htbp]
\centering
\includegraphics[width=14cm]{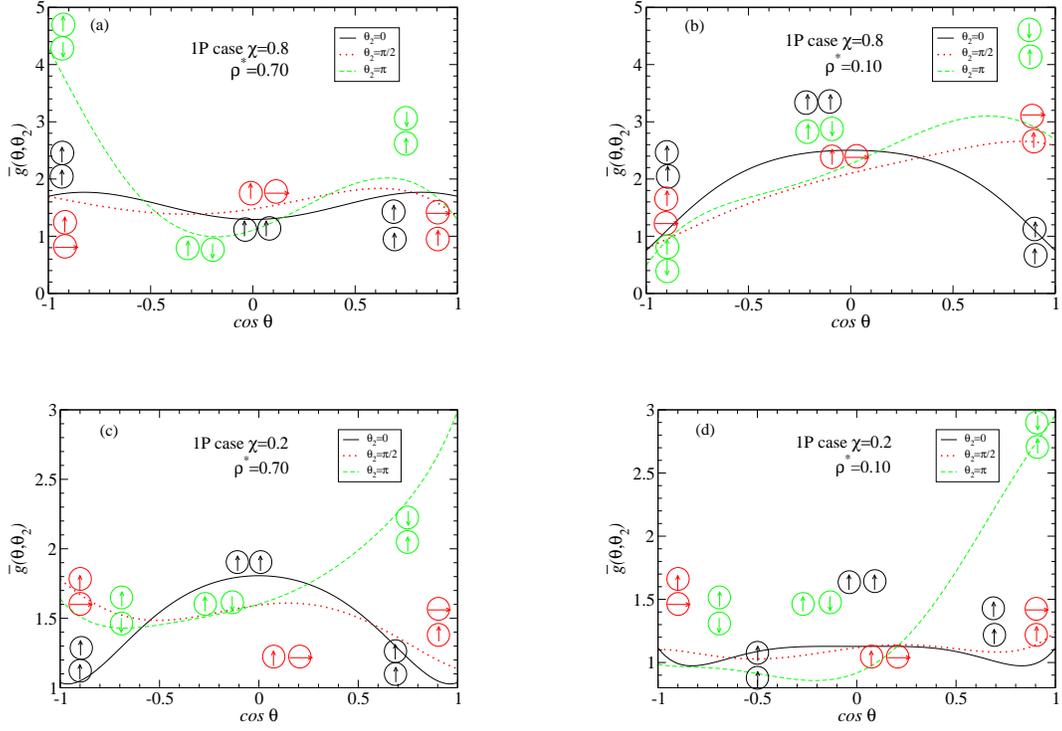}
\caption{Angular distribution $\bar{g}(\theta,\theta_2)$ for the one-patch model as a function of $\cos \theta$ for three different orientations of the patch on sphere $2$, given that sphere $1$ is fixed with patch along the $\hat{\mathbf{z}}$ axis. This is the one-patch counterpart of Fig. \ref{fig:fig7}. Results are reported for two different coverages, $\chi=0.8$ ((a) and (b)) and $\chi=0.2$ ((c) and (d)), and two different densities, $\rho^{*}=0.7$ ((a) and (c)) and $\rho^{*}=0.1$ ((b) and (d)), at the same temperature $T^{*}=1.0$ and square-well width $\lambda=1.5$. The colored arrows are cartoons of the orientations of the sphere $2$ patch, corresponding to $\theta_2=0,\pi/2,\pi$. 
\label{fig:fig10}}
\end{figure}

For the single patch with $\chi=0.2$ coverage (Fig. \ref{fig:fig10}, bottom panels (c) and (d)) parallel patches ($\theta_2=0$) are more likely in the perpendicular direction ($\theta\approx \pi/2$ or $\cos \theta \approx 0$), whereas antiparallel patches ($\theta_2=\pi$) are more likely in the forward direction ($\cos \theta \approx  1$). The case of perpendicular patches ($\theta_2=\pi/2$) are conversely more or less equally distributed along the whole angular region  $0\le \theta \le \pi$. There is no qualitative difference between the situation of high ($\rho^{*}=0.7$) and low ($\rho^{*}=0.1$) densities as shown by the contrast between the bottom left (c) and right (d) panels. Note that the result significantly contrasts with the corresponding results of the two-patches case depicted in Fig. \ref{fig:fig7}. Consider now the opposite situation of very large coverage ($\chi=0.8$) (Fig. \ref{fig:fig10}, (a)) where there is a single well-defined peak for antiparallel orientations ($\theta_2=\pi$) in the backward direction ($\cos \theta \approx -1$). Again, this markedly differs from the two-patch case (Fig. \ref{fig:fig7}, top left panel (a)), where there is a triple peak for aligned patches ($\theta_2=0,\pi$) in the forward ($\cos \theta \approx 1$), perpendicular ($\cos \theta \approx 0$), and backward ($\cos \theta \approx -1$) orientations. This is a dense state point. Under diluted conditions, $\rho^{*}=0.1$, we find a qualitatively similar behavior as in the dense case, with antiparallel alignment in the forward direction (which cannot be physically distingushed from the backward one). Clearly the predominant antiparallel alignment is reflecting the tendency to micellization rather than polymerization which is built into the single patch symmetry.

It is also interesting to contrast the coefficients of rotational invariants for the one-patch case with those obtained in the two-patch counterpart in Fig. \ref{fig:fig7}. At variance with the two-patch case, here all coefficients are nonvanishing so that we consider the first nonvanishing instances, that is $h^{110}(r)$, $h^{112}(r)$, and $h^{220}(r)$, which are particularly useful as giving the projections over the important invariants. \cite{Weis93}

\begin{figure}[htbp]
\centering
\includegraphics[width=14cm]{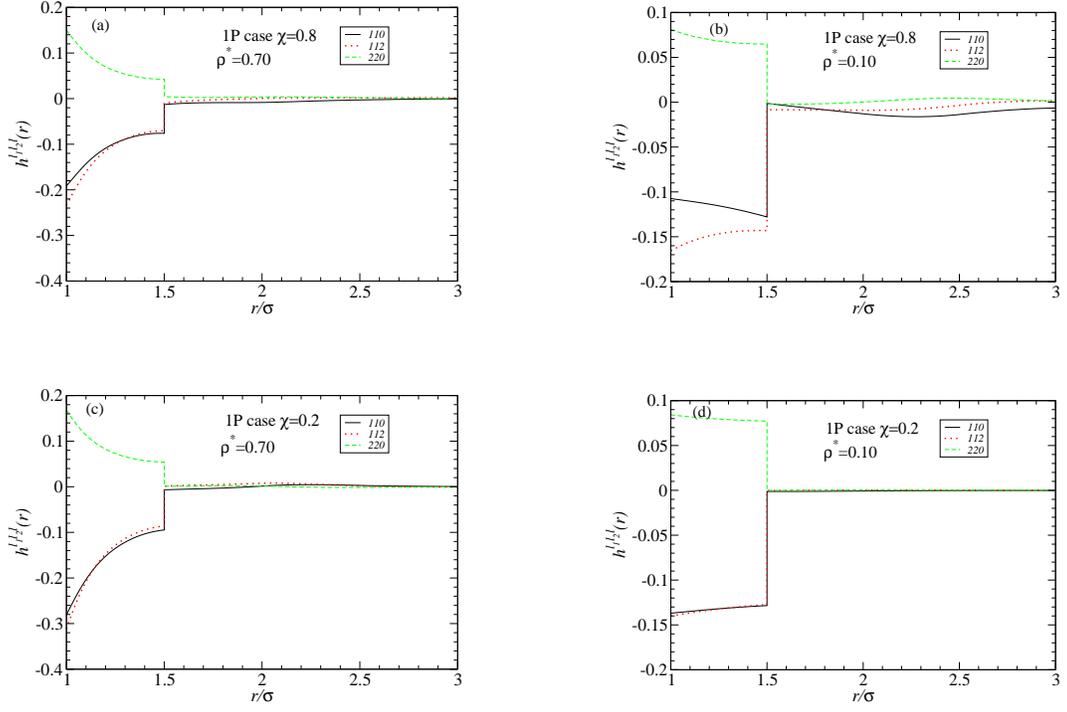}
\caption{Plot of rotational invariants $h^{110}(r)$, $h^{112}(r)$, and $h^{220}(r)$ for the one-patch model as functions of $r/\sigma$. 
Results are reported for two different coverages, $\chi=0.8$ ((a) and (b)) and $\chi=0.2$ ((c) and (d)), and two 
different densities, $\rho^{*}=0.7$ ((a) and (c)) and $\rho^{*}=0.1$ ((b) and (d)), at the same temperature $T^{*}=1.0$
and square-well width $\lambda=1.5$. Again, these are the same state points considered in Fig. \ref{fig:fig8}. 
\label{fig:fig11}}
\end{figure}

We have evaluated these coefficients for the same state points considered in the two-patch case in Fig. \ref{fig:fig11} for both dense or diluted conditions and small or large coverages. In contrast with the two-patch case, here the effect of coverage appears to be less significant, as can be inferred by inspection of the dense case $\rho^{*}=0.7$ (left panels (a) and (c) ). For the $h^{110}(r)$ case 
--- the projection coefficient along the ferroelectic invariant $\Delta(12)$ in Appendix \ref{app:appb} --- 
we find a negative correlation within the well both for $\chi=0.8$ (top left panel (a) ) and $\chi=0.8$ (bottom left panel (c) ), as expected from the tendency to form antiparallel alignments. Likewise, the projection  $h^{112}(r)$ along the dipolar invariant $D(\hat{\mathbf{n}}_1,\hat{\mathbf{n}}_2,\hat{\mathbf{r}}_{12})$ is found to be negative and numerically similar to $h^{110}(r)$ at both coverages, again indicating negative correlation to dipolar alignment. The only positive correlation is found for the $h^{220}(r)$ component, which does not distinguish between up and down symmetry, in qualitative agreement with the two-patch analogue.

This situation is replicated in the diluted case (right two panels (b) and (d) ) with different numerical values, thus indicating that these correlations are signatures of robust orientational trends induced by the the particular one-side symmetry of the single-patch potential.

\section{Conclusions and outlook}
\label{sec:conclusions}
We have performed a detailed study of a fluid whose particles interact via a two-patch Kern-Frenkel potential that attributes
a negative square-well energy whenever any two patches on the spheres are within a solid angle associated with a predefined coverage and
within a given distance given by the well width, and a simple hard-sphere repulsion otherwise. This model can be reckoned as a paradigm
of a unit system with incompatible elements ({\em e.g.}, hydrophobic and hydrophilic) that can self-assemble into different complex superstructures
depending on the parameters of the original unit ({\em e.g.}, coverage). We have exploited state-of-the-art numerical simulations (standard Metropolis,
Gibbs ensemble, and grand canonical Monte Carlo) coupled with RHNC integral equation theory following the approach outlined in previous work
on a single patch. \cite{Giacometti09b} On comparing RHNC integral equation with numerical simulations, we find the former to be quantitatively 
predictive in a large region of coverage, even close to the gas-liquid transition critical region, over a range of coverage which is significantly
larger than the single patch counterpart. The reason for this is attributed to the fact that RHNC uses the approximated hard-sphere bridge function, which retains spherical information, as a unique approximation throughout the entire calculation, a feature which works better for the more
symmetric two-patch case than the highly asymmetric one-patch Kern-Frenkel potential.

Having assessed the reliability of the RHNC integral-equation approach, we have fully exploited its capabilities in providing detailed angular information that is typically inaccessible to MC simulations, as already discussed in Ref. \onlinecite{Giacometti09b}. This has been done in two ways. First, by computing the orientational distribution probability of parallel and perpendicular alignment of patches within a spherical shell in the region $\sigma < r < \lambda \sigma$. This methodology is able to account for the erratic coverage dependence of the pair correlation function $g(12)$ by clearly discriminating between small and large coverages at all densities. The same approach also enlightens the characteristic symmetries of the patch distributions when the two-patch case result is contrasted with the one-patch analogue. Second, by computing the rotational-invariant coefficients that are the projections of $g(12)$ over rotational invariants. Again, this can discriminate between small and large coverages (at all densities) and single and double patches.

Our Monte Carlo results extend those originally obtained by Kern and Frenkel \cite{Kern03} for the two-patch case and can be contrasted with those of the corresponding single-patch counterpart \cite{Giacometti09b} and those obtained when the radial part of the potential is of the Baxter type. \cite{Fantoni07} The RHNC calculation presented here, along with the corresponding calculation carried out in our previous paper, \cite{Giacometti09b} together constitute the first attempt to apply a well-defined integral equation theory to such highly anisotropic potentials having sharp angular modulation.

An important outcome of our calculations is the clarification of the combined effect that size and distribution of the patches have on the
gas-liquid coexistence lines and critical parameters. The reduction of the bonding surface clearly decreases the critical temperature, an effect which can be related to the decrease in the bonding energy of the system. More interestingly, it also shows a suppression of the critical density, which can be interpreted along the same lines used in interpreting the valence dependence in patchy colloids. \cite{zacca1, Bianchi07} Indeed, the maximally bonded structures require lower and lower local densities on decreasing $\chi$. Interestingly, while in the single-bond-per-patch condition the evolution of the critical parameters on decreasing valence can be followed down to the limit  where clustering prevents phase separation, \cite{Sciortino09} in the model studied here crystallization pre-empts the observation of the liquid-gas separation for $\chi<0.3$.  Crystallization is here much more effective due to the analogy between the local fully-bonded configuration and the crystal structure. By contrast, crystallization is never observed for the one-patch case, where it has been shown that the lowest energy configuration is reached instead via the process of formation of large aggregates (micelles and vesicles) or via the formation of lamellar phases. \cite{Sciortino09} This difference highlights the important coupling between the orientational part of the potential and the possibility of forming extended fully-bonded structures. Our results indicate that, for a given coverage, in the two-patch case both the critical temperature and density are slightly higher then their corresponding one-patch counterparts, thus indicating that an increase of the valence favours the gas-liquid transition, in agreement with previous findings.

A final important consequence of our study concerns the limit of very small coverages that is particularly interesting. Indeed, it is possible to tune the structure of the system and control the topology of its ordered arrangement. By doing this we have observed a progression from the case where chains are stable (in the one-bond-per-patch limit) to the case where independent planes are found, evolving --- for slightly larger $\chi$ values --- into an ordered three-dimensional crystalline structure.  Each of these ordered structures is observed in a restricted range of $\chi$ values. This possibility of fine tuning the morphology  by controlling  the patterning of the particle surfaces may offer an interesting possibility for specific self-assembling structures.

An additional perspective of our work should be stressed. Several studies (see {\em e.g.} Ref. \onlinecite{Giacometti05} and references therein) have exploited spherically-symmetric potentials to mimic effective protein-protein interactions, especially in connection with protein crystallization. \cite{George94} This is clearly unrealistic for the majority of proteins where the distribution of hydrophobic surface groups is significantly irregular, a feature that can be captured, at the simplest level of description, by the model studied here. The specific location of the coexistence lines, such as those considered in the present study, have important consequences in the study of pathogenic events for sickle cell anemia \cite{Galkin02} and other human diseases. \cite{Eton90}

\section{Acknowledgements}
We thank Philip J. Camp and Enrique Lomba for useful suggestions.
FS acknowledges support from  NoE SoftComp NMP3-CT-2004-502235, ERC--226207--PATCHYCOLLOIDS  and
ITN-COMPLOIDS. AG acknowledges support from PRIN-COFIN 2007B57EAB(2008/2009). 

\appendix
\section{Angular properties of $g(12)$ in a general frame}
\label{app:appa}

The expansion in spherical harmonics $Y_{lm}(\omega)$ of $g(12)$ in an arbitrary space frame reads \cite{Note}
\begin{eqnarray}
\label{appa:eq1}
g\left(12\right) &=& 4 \pi \sum_{l_{1},l_{2}=0}^{\infty} \sum_{l=\vert l_1 - l_2 \vert}^{l_1+l_2} g\left(r;l_{1} l_{2} l\right)
\psi^{l_{1} l_{2} l}\left(\omega_1 \omega_2 \Omega\right),
\end{eqnarray}
where we have introduced the rotational invariants \cite{Gray84}
\begin{eqnarray}
\label{appa:eq2}
\psi^{l_{1} l_{2} l}\left(\omega_1 \omega_2 \Omega\right)&=& \sum_{m_{1}=-l_{1}}^{+l_{1}}  \sum_{m_{2}=-l_{2}}^{+l_{2}}
C\left(l_{1} l_{2} l; m_{1} m_{2} m_{1}+m_{2} \right)
Y_{l_{1}m_{1}}\left(\omega_{1}\right) Y_{l_{2}m_{2}}\left(\omega_{2}\right)
Y_{l, m_{1}+m_{2}}^{*}\left(\Omega\right).
\end{eqnarray}
Note that $g\left(r;l_{1} l_{2} l\right)$ coincides with $g^{l_{1} l_{2} l}(r)$ up to a normalization constant (see Appendix \ref{app:appb}).

We are free to set the origin of the coordinate frame at the center of particle $1$ and choose its orientation so that $\hat{\mathbf{z}}=\hat{\mathbf{n}}_1$ without loss of generality. We first note that
\begin{eqnarray}
\label{appa:eq4}
Y_{l_{1} m_{1}} \left(\theta_1=0, \varphi_1\right)&=& \left( \frac{2l_1+1}{4\pi} \right)^{1/2} \delta_{m_10}.
\end{eqnarray}
Clearly, the orientation of the $x,y$ axes is then irrelevant, so we may integrate out the angles $\varphi_2$ and  $\varphi$; this leads to the average $\langle g(12) \rangle_{\varphi_{2} \varphi}$, where we note that 
\begin{eqnarray}
\label{appa:eq3}
\left \langle Y_{l_2m_2}\left(\theta_2, \varphi_2\right) \right \rangle_{\varphi_{2}} &=& 
\left(-1\right)^{m_2} \left[ \frac{2l_2+1}{4\pi}
\frac{\left(l_2-m_2\right)!}{\left(l_2+m_2\right)!} \right]^{1/2} P_{l_{2} m_{2}}\left(\cos \theta_2\right) \frac{1}{2\pi} \int_{0}^{2\pi} d\varphi_2 
e^{\mathrm{i} m_2 \varphi_2} \\ \nonumber 
&=& \left( \frac{2l_2+1}{4\pi} \right)^{1/2} P_{l_{2}}\left(\cos \theta_2\right) \delta_{m_20}
\end{eqnarray}
and similarly for $\left \langle Y_{l m_2}^*\left(\theta, \varphi \right) \right \rangle_{\varphi}$. Here the $P_{l0}(x)=P_{l}(x)$ are the usual Legendre polynomials. We have then from Eqs. (\ref{appa:eq1}) and (\ref{appa:eq2}) that
\begin{eqnarray}
\left \langle g\left(12\right) \right \rangle_{\varphi_2 \varphi}&=& 
\sum_{l_{1},l_{2},l} g \left( r;l_{1} l_{2} l \right) 
\left[ \frac{\left(2l_1+1\right) \left(2l_2+1\right) \left(2l+1\right)}{4\pi} \right]^{1/2} C\left(l_{1} l_{2} l; 0 0 0 \right) P_{l_{2}}\left(\cos \theta_2\right) 
P_{l}\left(\cos \theta\right).
\label{appa:eq5}
\end{eqnarray}
We are interested in the angular behavior within the well, $\sigma \le r \le \lambda \sigma$, so we finally integrate over the radial variable $r$ and define
\begin{eqnarray}
\label{appa:eq6}
\bar{g}\left(l_1 l_2 l\right) &=& \frac{1}{(\lambda-1) \sigma} \int_{\sigma}^{\lambda \sigma} dr g\left(r; l_{1} l_{2} l\right), \\
\bar{g}\left(\theta,\theta_2\right) &=& \frac{1}{(\lambda-1) \sigma} \int_{\sigma}^{\lambda \sigma} dr \left \langle g\left(12\right) \right \rangle_{\varphi_2 \varphi}. 
\end{eqnarray}
The result is then a function of the polar coordinate $\theta$ of $\hat{\mathbf{r}}_{12}$ and the polar orientation of the second patch, $\theta_2$; that is,
\begin{eqnarray}
\label{appa:eq7}
\bar{g}\left(\theta,\theta_2\right) &=& 
\sum_{l_{1},l_{2},l} \bar{g}\left(l_1 l_2 l\right)
\left[ \frac{\left(2l_1+1\right) \left(2l_2+1\right) \left(2l+1\right)}{4\pi} \right]^{1/2} C\left(l_{1} l_{2} l; 0 0 0 \right) P_{l_{2}}\left(\cos \theta_2\right) 
P_{l}\left(\cos \theta\right),
\end{eqnarray}
given that the $z$ axis is aligned with the patches of particle 1.

\section{Coefficients $h^{l_{1} l_{2} l}(r)$ of rotational invariants}
\label{app:appb}

In this Appendix we consider the coefficients $g^{l_{1} l_{2} l}(r)$ of rotational invariants that have proven to be particularly useful in discriminating among different orientational behaviors. In numerical simulations they are defined as follows (see {\em e.g.} Ref. \onlinecite{Weis93}),
\begin{eqnarray}
\label{appb:eq1}
g^{l_{1} l_{2} l}\left(r\right) &=& \frac{1}{N 4 \pi \rho r^2} \left \langle \sum_{i\ne j} \delta\left(r-r_{ij}\right) \Delta^{l_{1} l_{2} l} (12) \right \rangle,  
\end{eqnarray}
where the $\Delta^{l_{1} l_{2} l} (12)$ are rotational invariants. Explicit expressions for the first few are \cite{Stell81} 
\begin{eqnarray}
\label{appb:eq2}
\Delta^{000}\left(12\right) &=&1, \\ \nonumber
\Delta^{110}\left(12\right) &=& 3\Delta\left(12\right)=3\,\hat{\mathbf{n}}_1 \cdot \hat{\mathbf{n}}_2, \\ \nonumber
\Delta^{112}\left(12\right)&=& \frac{3}{2} D\left(12\right)= \frac{3}{2} \left[3 \left(\hat{\mathbf{n}}_1 \cdot \hat{\mathbf{r}}_{12} \right)
\left(\hat{\mathbf{n}}_2 \cdot \hat{\mathbf{r}}_{12} \right) - \hat{\mathbf{n}}_1 \cdot \hat{\mathbf{n}}_2 \right], \\ \nonumber
\Delta^{220}\left(12 \right)&=& \frac{5}{2} E\left(12\right)=\frac{5}{2} \left[3\left( \hat{\mathbf{n}}_1 \cdot \hat{\mathbf{n}}_2\right)^2 -1\right].
\end{eqnarray} 
Other expressions can be found in Ref. \onlinecite{Stell81}.

We note that the first expression in Eqs. (\ref{appb:eq2}) yields $g^{000}(r)$, which coincides with the radial distribution function
$g_{000}(r)=\langle g(\mathbf{r},\omega_1, \omega_2) \rangle_{\omega_1 \omega_2}$. Here we have $h^{000}(r)=g^{000}(r)-1$; in all other cases  $h^{l_{1} l_{2} l}(r)=g^{l_{1} l_{2} l}(r)$. Some of the coefficients have particularly interesting physical interpretations:   
the term $h^{110}(r)$ is the coefficient of ferroelectric correlation, the term $h^{112}(r)$ the coefficient of dipolar correlation, 
the term $h^{220}(r)$ the coefficient of nematic correlation, and so on.

It might be useful to show how these general expressions (typically computed in Monte Carlo simulations) connect with the corresponding ones
typically evaluated in an integral equation approach. We do this for the representative case of $g^{112}(r)$, the others being similar.

We define $g^{l_1l_2l}(r) \propto g\left(r;l_{1} l_{2} l\right)$, where the proportionality constant is obtained through a particular prescription
to be further elaborated below. The projections $g(r;l_1 l_2 l)$ of $g(12)$ on the rotational invariants $\psi^{l_{1} l_{2} l}(\omega_1 \omega_2 \Omega)$ as defined by Eq. (\ref{appa:eq1}) are related to the values $g_{l_{1} l_{2} m}(r)$ in the axial frame by \cite{Gray84}
\begin{eqnarray}
\label{appb:eq3}
g\left(r;l_{1} l_{2} l\right) &=& \left( \frac{4\pi}{2 l+1} \right)^{1/2} 
\sum_m C\left( l_1 l_2 l; m \bar{m} 0 \right) g_{l_{1} l_{2} m} \left(r\right),
\end{eqnarray}
where $\bar{m} \equiv -m$. 

Consider as a representative example the quantity $\Delta^{112}(12)$ defined in Eqs. (\ref{appb:eq2}) and note that in Eq. (\ref{appb:eq1}) one has
\begin{eqnarray}
\left \langle \sum_{i\ne j=1}^N\delta\left(r-r_{ij}\right) D\left(ij\right) \right \rangle &=& \frac{N \rho r^2}{4 \pi} \int d \omega_1 d \omega_2 \, g\left(r,\omega_1, \omega_2 \right)D\left(12\right).
\label{appb:eq4}
\end{eqnarray}
Upon choosing the molecular frame ($\hat{\mathbf{z}}\equiv \hat{\mathbf{r}}_{12}$), one finds
\begin{eqnarray}
\label{appb:eq5}
D\left(12\right) &=& 2 \cos \theta_1 \cos \theta_2 - \sin \theta_1 \sin \theta_2 \cos \left(\varphi_1 -\varphi_2\right)\\ \nonumber
&=&\frac{4 \pi}{3} \left[ 2 Y_{10}\left(\omega_1\right) Y_{10}\left(\omega_2\right) + Y_{11}\left(\omega_1\right)
Y_{1-1} \left(\omega_2\right)+Y_{1-1}\left(\omega_1\right) Y_{11} \left(\omega_2\right) \right],
\end{eqnarray}
where the $Y_{lm}(\omega)$ are spherical harmonics. \cite{Gray84}

The expansion (\ref{appa:eq1}) in a molecular frame reduces to
\begin{eqnarray}
\label{appb:eq6}
g\left(r,\omega_1, \omega_1\right) &=& 4 \pi \sum_{l_1 l_2 m} g_{l_1 l_2 m} \left(r\right) Y_{l_1 m} \left(\omega_1\right) 
Y_{l_2 \bar{m}} \left(\omega_2\right).
\end{eqnarray}
Using Eqs. (\ref{appb:eq5}), (\ref{appb:eq6}), and the orthogonality relations for spherical harmonics, \cite{Gray84} one finds easily that
\begin{eqnarray}
\label{appb:eq7}
\frac{1}{(4\pi)^2} \int d \omega_1 d \omega_2 \, g\left(r,\omega_1 \omega_2 \right) D\left(12\right) &=& \frac{1}{3} \left[2
g_{110}\left(r\right) + g_{111}\left(r\right) + g_{11-1} \left(r \right) \right],
\end{eqnarray}
so that, combining with $h^{112}(r)$ given in Eqs. (\ref{appb:eq1}) and (\ref{appb:eq2}) and using the symmetry property $g_{11-1}(r)=g_{111}(r)$, one finds
\begin{eqnarray}
\label{appb:eq8}
h^{112}\left(r\right) &=& g_{110}\left(r\right)+ g_{111}\left(r\right).
\end{eqnarray}

We can work out the first few projections by using tabulated values for the Clebsch-Gordan coefficients $C\left( l_1 l_2 l; m \bar{m} 0 \right)$. 
Using the symmetry properties of the Clebsch-Gordan coefficients and the $g_{l_{1} l_{2} m_{1}}(r)$, one finds
\begin{eqnarray}
g\left(r;000\right)&=& (4 \pi)^{1/2} g_{000}\left(r\right), \\ \nonumber
g\left(r;110\right)&=&-\left( 4\pi/3 \right)^{1/2} \left[ g_{110}\left(r\right)-2 g_{111}\left(r\right) \right], \\ \nonumber
g\left(r;112\right)&=& \left( 8\pi/15 \right)^{1/2} \left[ g_{110}\left(r\right) +g_{111}\left(r\right) \right], \\ \nonumber
g\left(r;220\right)&=& \left( 4\pi/5 \right)^{1/2} \left[g_{220}\left(r\right)-2 g_{221}\left(r\right)+2g_{222}\left(r\right) \right], \\ \nonumber
g\left(r;222\right)&=& -\left( 8\pi/35 \right)^{1/2}\left[g_{220}\left(r\right)
-g_{221}\left(r\right)-2 g_{222}\left(r\right) \right], \\ \nonumber
g\left(r;224\right)&=& \left( 8\pi/35 \right)^{1/2} \left[g_{220}\left(r\right)+\frac{4}{3} g_{221}\left(r\right)+ \frac{1}{3}g_{222}\left(r\right)  \right], \\ \nonumber 
g\left(r;011\right)&=& \left( 4\pi/3 \right)^{1/2} g_{010}\left(r\right) = -g\left(r;101\right), \\ \nonumber
g\left(r;022\right)&=& \left( 4\pi/5 \right)^{1/2} g_{020}\left(r\right)=g\left(r;202\right), \\ \nonumber
g\left(r;121\right)&=& -\left( 8\pi/15 \right)^{1/2} \left[g_{120}\left(r\right)-\sqrt{3}\, g_{121}\left(r\right) \right]=-g\left(r;211\right), \\ \nonumber
g\left(r;123\right)&=& \left( 12\pi/35 \right)^{1/2}\left[g_{120}\left(r\right)+
\frac{2}{\sqrt{3}} g_{121}\left(r\right) \right]=-g\left(r;213\right). 
\label{appb:eq9}
\end{eqnarray}
As anticipated above, we now fix the proportionality constant in $g^{l_1l_2l}(r) \propto g\left(r;l_{1} l_{2} l\right)$ by dividing out the leading constants above so that the coefficient of $g_{l_1l_20}(r)$ is unity; {\em e.g.}, $g^{220}(r)=g_{220}\left(r\right)-2 g_{221}\left(r\right)+2g_{222}\left(r\right)$. 


\bibliographystyle{apsrev}

\begin{thebibliography}{99}
\bibitem{Note} Some authors omit the $4\pi$ prefactor in this expansion. For instance, our $g(r;l_1 l_2 l)$ are related to
the $g(l_1 l_2 l;r)$ appearing in Ref. \onlinecite{Gray84} by $g(l_1 l_2 l;r)=4 \pi g(r;l_1 l_2 l)$.
\end{thebibliography}

\clearpage
\begin{table}
\begin{center}
\begin{tabular}{lccccc}
\hline 
$\chi$ & $\rho^*_c (\rm{MC})$ & $\rho^*_c (\rm{RHNC})$ & $T^*_c (\rm{MC})$ &$T^*_c (\rm{RHNC})$  & $z_c (\rm{MC}) $\\
\hline
1.0 & 0.312&0.317 &1.22 & 1.23 & 0.0526\\
0.9 & 0.309&0.296 &1.05 & 1.10 & 0.0516\\
0.8 & 0.303&0.277 &0.883 &0.949 & 0.0487\\
0.7 & 0.287&0.262 &0.714 & 0.750 & 0.0432\\
0.6 & 0.262&0.254 &0.555 & 0.562 & 0.0350 \\
0.5 & 0.234&      &0.423 &  &0.0202\\
0.4 & 0.206&      &0.333 &  & 0.0202\\
0.3 & 0.175&      &0.257 & &0.0155 \\
\hline 
\end{tabular}
\caption{Comparison of the estimated location of critical points from MC and RHNC data (see text).  In the last column  $z_c \equiv e^{\beta \mu}$ 
is the critical activity.
\label{tab:tab1}
}
\end{center}
\end{table}
\begin{table}
\begin{center}
\begin{tabular}{lccccccccccccccc}
\hline
$\chi$& &$U/N\epsilon$& &$\beta F_{\text{ex}}/N$& &$\beta \mu$& &$\beta P/\rho$&
&$\beta \left(\partial P/\partial \rho\right)_T$& &$\sigma_0/\sigma$& &$\bar{z}$ \\
\hline
$1.0$ && $-5.46$     &&$-2.56$  && $-3.28$      &&$0.64$   &&  $10.33$ && $1.031$ &&$10.92$   \\
$0.9$ && $-4.88$     &&$-1.81$  && $-1.79$      &&$1.37$   &&  $11.26$ && $1.026$ &&$ 9.76$   \\
$0.8$ && $-4.10$     &&$-0.98$  && $-0.10$      &&$2.24$   &&  $12.39$ && $1.020$ &&$ 8.20$   \\
$0.7$ && $-3.25$     &&$-0.19$  && $ 1.53$      &&$3.08$   &&  $13.54$ && $1.014$ &&$ 6.50$   \\
$0.6$ && $-2.49$     &&$ 0.49$  && $ 2.88$      &&$3.75$   &&  $14.48$ && $1.010$ &&$ 5.00$   \\
$0.5$ && $-1.88$     &&$ 1.05$  && $ 3.94$      &&$4.25$   &&  $15.21$ && $1.007$ &&$ 3.76$   \\
$0.4$ && $-1.26$     &&$ 1.61$  && $ 5.00$      &&$4.74$   &&  $15.95$ && $1.005$ &&$ 2.51$   \\
$0.3$ && $-0.80$     &&$ 2.00$  && $ 5.75$      &&$5.10$   &&  $16.51$ && $1.003$ &&$ 1.60$   \\
$0.2$ && $-0.39$     &&$ 2.33$  && $ 6.39$      &&$5.42$   &&  $17.00$ && $1.001$ &&$ 0.79$   \\
$0.1$ && $-0.11$     &&$ 2.54$  && $ 6.79$      &&$5.62$   &&  $17.31$ && $1.000$ &&$ 0.23$    \\
$0.0$ && $ 0.00$     &&$ 2.61$  && $ 6.95$      &&$5.69$   &&  $17.43$ && $1.000$ &&$ 0.00$    \\
\hline
\end{tabular}
\caption[]{Values of reduced internal and excess free energies, chemical potential, pressure, and inverse compressibility as a function of the coverage $\chi$ for a fixed state point, $\rho^{*}=0.7$ and $T^{*}=1.0$. The last two columns report the reference HS diameter $\sigma_0$ (in units of $\sigma$) and the average coordination number $\bar{z}$. Expected errors are in the last digits.
\label{tab:tab2}
}
\end{center}
\end{table}

\clearpage
\clearpage
\clearpage

\clearpage
\clearpage
\clearpage
\clearpage
\clearpage
\clearpage
\clearpage
\clearpage
\end{document}